\documentclass[]{aa}
\usepackage{graphicx}
\usepackage{txfonts}
\usepackage{xspace}
\usepackage{comment}
\usepackage{amsmath}
\usepackage{gensymb}
\usepackage[colorlinks, citecolor={blue}]{hyperref}
\usepackage{multirow}
\usepackage{lscape}
\usepackage{longtable, supertabular}

\newcommand{\chandra}{\textit{Chandra}\xspace}
\newcommand{\xmm}{\textit{XMM-Newton}\xspace}
\newcommand{\nustar}{\textit{NuSTAR}\xspace}
\newcommand{\suzaku}{\textit{Suzaku}\xspace}
\newcommand{\bat}{\textit{Swift}/BAT\xspace}

\newcommand{\fluxcgs}{erg cm$^{-2}$ s$^{-1}$\xspace}
\newcommand{\lumcgs}{erg s$^{-1}$\xspace}

\newcommand{\nh}{$N_{\rm H}$\xspace}

\begin{document} 

   \title{A comprehensive X-ray view of the active nucleus in NGC 4258}

   \author{Alberto Masini\inst{1,2}
          \and
          J. V. Wijesekera\inst{3}
          \and
          Annalisa Celotti\inst{1,4,5,6}
          \and
          Peter G. Boorman\inst{7,8}
          }

   \institute{SISSA - International School for Advanced Studies, Via Bonomea 265, I-34151 Trieste, Italy 
   \and 
   INAF - Osservatorio di Astrofisica e Scienza dello Spazio di Bologna, Via Gobetti 93/3, I-40129 Bologna, Italy 
   \and
   Department of Physics and Astronomy, University of Padova, Via Francesco Marzolo 8, I-35121 Padova, Italy
   \and
   INAF - Osservatorio Astronomico di Brera, Via Bianchi 46, I-23807 Merate, Italy
   \and
   IFPU - Institute for Fundamental Physics of the Universe, Via Beirut 2, I-34151 Trieste, Italy
   \and
   INFN - National Institute for Nuclear Physics, Via Valerio 2, I-34127 Trieste, Italy
   \and
   Astronomical Institute, Academy of Sciences, Bo\v{c}n\'i II 1401, CZ-14100 Prague, Czech Republic
   \and
   Department of Physics \& Astronomy, University of Southampton, Southampton SO17 1BJ, UK
  }
   
   \date{}

  \abstract
   {The presence of water masers orbiting around the active galactic nucleus (AGN) in NGC 4258, one of the most studied extragalactic objects, has been crucial in developing a detailed picture of its nuclear environment. Despite this, its accretion rate and bolometric luminosity are still matter of debate, as there are indications that NGC 4258 may host a genuine radiatively inefficient accretion flow (RIAF).}
   {In this respect, we present a detailed broadband X-ray spectrum of NGC 4258, with the goal of precisely measuring the coronal luminosity and accretion flow properties of the AGN, and track any possible variation across two decades of observations.}
   {We collect archival \xmm, \chandra, \bat and \nustar spectroscopic observations spanning 15 years, and fit them with a suite of state of the art models, including a warped disk model which is suspected to provide the well known obscuration observed in the X-rays. We complement this information with archival results from the literature.}
   {Clear spectral variability is observed among the different epochs. The obscuring column density shows possibly periodic fluctuations on a timescale of 10 years, while the intrinsic luminosity displays a long term decrease of a factor of three in a time span of 15 years (from $L_{2-10~\text{keV}}\sim 10^{41}$ \lumcgs in the early 2000s, to $L_{2-10~\text{keV}}\sim 3\times 10^{40}$ \lumcgs in 2016). The average absorption-corrected X-ray luminosity $L_{2-10~\text{keV}}$, combined with archival determinations of the bolometric luminosity, implies a bolometric correction $k_{\rm bol} \sim 20$, intriguingly typical for Seyferts powered by accretion through  geometrically thin, radiatively efficient disks. Moreover, the X-ray photon index $\Gamma$ is consistent with the typical value of the broader AGN population. However, the accretion rate in Eddington units is very low, well within the expected RIAF regime.}
   {Our results suggest that NGC 4258 is a genuinely low-luminosity Seyfert II, with no strong indications in its X-ray emission for a hot, RIAF-like accretion flow.}

   \keywords{galaxies: active}

   \maketitle


\section{Introduction}
The nearby spiral galaxy NGC 4258 hosts one of the closest active galactic nuclei (AGN), at a distance of $7.58 \pm 0.11$ Mpc \citep{reid19}. However, its historical importance goes beyond its mere distance. Since the discovery of nuclear megamaser emission at 22 GHz from water vapour molecules \citep{claussen84}, NGC 4258 soon became the cleanest evidence for the existence of extragalactic supermassive black holes (SMBHs). The mapping and temporal monitoring of the masers through Very Long Baseline Interferometry (VLBI) revealed a sub-pc molecular, dusty disk in Keplerian rotation around a central mass $M = 4 \times 10^7 M_\odot$ \citep{nakai93, miyoshi95}. Moreover, NGC 4258 has since been an anchor galaxy to calibrate the distance ladder and to measure the expansion rate of the Universe $H_0$ \citep{pesce20}. Nowadays, water masers tracing edge-on molecular disks around AGN -- called disk megamasers -- have been (more or less securely) detected in a couple dozen of galaxies, and allow to measure both the most precise extragalactic SMBH masses to date \citep[e.g.,][]{kuo11}, and geometric distances \citep[e.g.,][]{kuo15, reid19}. NGC 4258 has historically been considered as the archetype for the class of disk megamasers, although today it is well accepted to be a somewhat anomalous source with respect to the others \citep[e.g., for its low obscuration along the line of sight; see][]{masini16}. 

After the discovery of water maser spots in the nucleus of NGC 4258, slight, albeit significant, deviations from a pure Keplerian rotation  were then noticed \citep{herrnstein05}, indicating that a simple flat, geometrically thin molecular disk was a too simplistic model to explain their dynamics. By introducing both a position angle and inclination warps, \citet{herrnstein05} were instead able to reconcile the maser spots kinematics with a pure Keplerian rotation. Furthermore, in this model the projected clustering of the systemic masers was naturally explained by their confinement in the bottom of the "bowl" caused by the inclination warp. A natural implication of a warped maser disk was that it would rise in front of the observer, hiding the central engine and possibly providing the observed moderate absorption in the X-ray spectrum \citep{fruscione05}, first noticed more than 25 years ago \citep{makishima94}. \citet{herrnstein05} were able to put a constraint on the radial distance at which the warp crosses the line of sight, by comparing their warp model with the radius at which the X-ray irradiation would cause a transition from molecular to atomic gas in the disk \citep{neufeldmaloney95}. The transition radius was found to be at $\sim 0.28$ pc from the nucleus, and was also used to constrain the accretion rate through the maser disk assuming a steady state, geometrically thin and optically thick accretion disk \citep{shakurasunyaev73}. 
\par Also, NGC 4258 is a very interesting AGN because its particularly low bolometric luminosity in Eddington units\footnote{The bolometric luminosity of NGC 4258 has been historically estimated in the literature through SED fitting \citep{lasota96, yuan02, wu13}, and has been found to be around $\sim10^{-4}$ times the Eddington luminosity. The optical/UV data employed by those works are either upper limits from continuum observations, or those by \citet{wilkes95}, who directly detected the nucleus of NGC 4258 in polarized light at 5500 \AA.} ($L_{\rm Bol}/L_{\rm Edd} \sim 10^{-4}$) could be either explained by a very low accretion rate, or by invoking a radiatively inefficient accretion flow \citep[RIAF; see, e.g.,][and references therein]{narayanyi95}. The acronym RIAF generally refers to  physically different accretion flows, such as advection-dominated flows \citep[ADAF,][]{narayanyi95} or adiabatic inflow-outflow solutions \citep[ADIOS,][]{blandfordbegelman99}. In these models, the inner regions of the accretion disk do not follow the usually assumed \citet{shakurasunyaev73} one. Therefore, a detailed study of this source offers a unique opportunity to explore the inner fraction of a parsec of an under-luminous AGN.
\par Several previous studies have investigated the nature of the accretion flow in NGC 4258, both through its broadband spectral energy distribution (SED) and detailed X-ray spectroscopy. After the discovery of the sub-pc maser disk, for instance, \citet{neufeldmaloney95} derived an accretion rate $\dot{M} = 7\times 10^{-5}\alpha$ \citep[where $\alpha$ is the standard viscosity parameter of][]{shakurasunyaev73}, by assuming a viscous accretion disk whose midplane is traced by the masers, and obliquely illuminated by a central X-ray source. By combining the low X-ray luminosity with such an accretion rate and assuming a bolometric correction typical of Seyfert galaxies \citep{mushotzky93}, \citet{neufeldmaloney95} derived a ``standard'' radiative efficiency of order $\eta \sim 0.1$, thus not requiring radiatively inefficient accretion to explain the low luminosity of the AGN. On the other hand, \citet{lasota96} and \citet{gammie99} fitted the SED of the source with an ADAF with a much higher accretion rate ($\dot{M} \sim 10^{-2}$). However, \citet{yuan02} later demonstrated that a ``classical'' ADAF-like accretion flow was not able to account for the nuclear IR data, in particular the steep power-law shape, indicative of non-thermal emission \citep{chary00}. Moreover \citet{fiore01}, analyzing BeppoSAX data, found that pure bremsstrahlung emission \citep[as expected from an ADAF,][]{narayanyi95} is ruled out by the X-ray data. Rather, the X-ray spectrum of NGC 4258 resembles an average Seyfert-like spectrum, being successfully described by the combination of a power-law modified at low energy by photoelectric absorption, and soft X-ray emission due to the extended (kpc-scale) structures connected to the so-called twisted anomalous arms \citep[and references therein]{wilson01}. \citet{yuan02} suggested that a composite model, in which an outer thin disk transitions to an inner RIAF, which then transfers energy to a relativistic jet, was able to account for the radio to X-ray SED of NGC 4258. In this model, the X-ray emission would arise from Comptonization by hot electrons at the base of the jet. This picture was broadly supported by \citet{herrnstein05}, who inferred a consistently low ($\sim 10^{-4}\alpha$ $M_\odot$ yr$^{-1}$) accretion rate. The absence of a broad Fe K$\alpha$ line, paired with the detection of a weak, narrow and rapidly variable component \citep{reynolds09}, further supported the view that the inner regions of NGC 4258 might deviate significantly from a canonical radiatively efficient disk. Later on, \citet{wu13} fit again the broadband nuclear SED of NGC 4258 with a composite model (inner RIAF + outer truncated thin disk + a jet), suggesting that the SED can be reproduced by a combination of the three components, and placing a constraint over the spin parameter of the SMBH ($a = 0.7 \pm 0.2$).

\par The goal of this work is to comprehensively review the X-ray properties of NGC 4258, to possibly shed new light on its accretion flow and long term ($\sim 20$ years) evolution. To this aim, we take advantage of the wealth of available data: we re-analyze with state of the art models archival (from $\sim 2000-2016$) spectroscopic observations of the \chandra X-ray telescope \citep{weisskopf00}, the \xmm observatory \citep{jansen01}, the \bat telescope \citep{gehrels04, barthelmy05}, and \nustar \citep{harrison13}. The obtained results are then complemented with others from the literature (from $\sim 1993-2000$), to obtain a complete and thorough X-ray view of a nearby under-luminous AGN spanning 23 years of observations.

\par The paper is structured as follows: in Section \ref{sec:datareduction}, both the observations used in this work and their associated data reduction are presented. Section \ref{sec:offnuc} discusses the contribution to the \xmm and \nustar fluxes of the non-AGN components, i.e. the large-scale plasma emission and the off-nuclear point source. The broadband spectral analysis is conducted in Section \ref{sec:analysis}, with the setup presented in \ref{sec:setup} and results discussed in Section \ref{sec:results} and \ref{sec:context}.
Both the short and long term variability are discussed in Section \ref{sec:variability}. The discussion of the general results obtained is found in Section \ref{sec:discussion}, while our conclusions are drawn in Section \ref{sec:conclusions}. Tables with spectral analysis results are reported in Appendix \ref{sec:appendix}.

\par Throughout the paper, uncertainties are given at the 90\% confidence level, unless otherwise stated, and no cosmology is assumed, given the adopted geometric distance \citep{reid19}.

\begin{table}
\caption{Log of the observations analyzed in this work.}
\label{tab:info}
\centering                                    
\begin{tabular}{l c c c}          
\hline\hline                  
Telescope & ObsID & Date & Exp (ks) \\ 
\hline
  \multirow{4}{5em}{\chandra} & 349  & 08-Mar-2000 & 2.8 \\ 
                              & 350  & 17-Apr-2000 & 14.2 \\
                              & 1618 & 28-May-2001 & 21.3 \\
                              & 2340 & 29-May-2001 & 7.6 \\
 \hline                             
  \multirow{8}{5em}{\xmm} & 0110920101 & 08-Dec-2000 & 13.3/20.1 \\
                          & 0059140101 & 06-May-2001 & 7.8/12.0\\
                          & 0059140201 & 17-Jun-2001 & 2.7/8.4\\
                          & 0059140401 & 17-Dec-2001 & 0.3/3.3\\
                          & 0059140901 & 22-May-2002 & 9.8/13.6\\
                          & 0400560301 & 17-Nov-2006 & 45.3/57.2\\ 
 \hline
  \multirow{2}{5em}{\nustar}  & 60101046002 & 16-Nov-2015 & 54.8 \\ 
                              & 60101046004 & 10-Jan-2016 & 103.6 \\
\hline    
\bat & $-$ & 105 months & $1.85 \times 10^5$ \\ 
\hline 
\end{tabular}
\tablefoot{Exposure times relative to the \xmm observations are the final, cleaned exposure times of each spectrum, for the PN and MOS cameras, respectively.}
\end{table}

\section{Observations and Data reduction}\label{sec:datareduction}

\begin{figure*}
   \centering
   \includegraphics[width=0.95\textwidth]{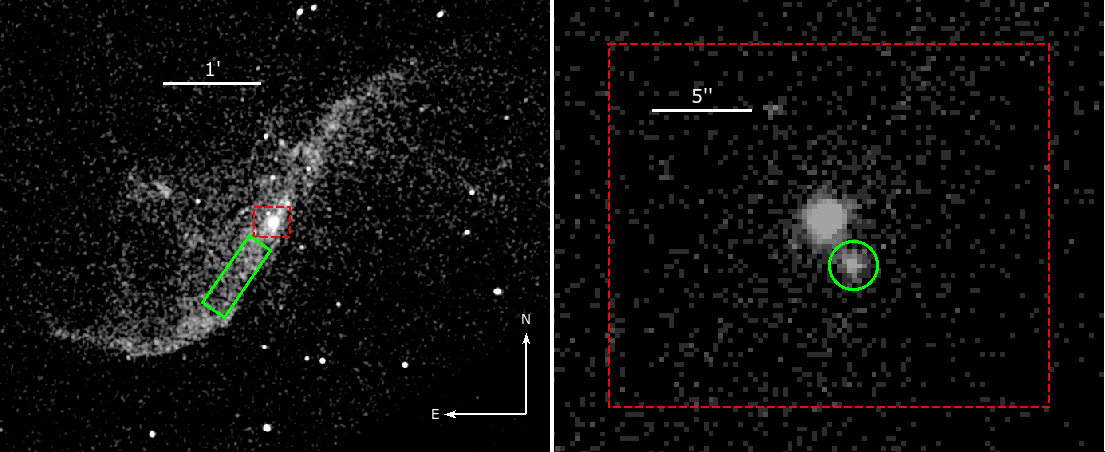}
   \caption{{\it Left.} \chandra ACIS-S $0.5-7.0$ keV image of NGC 4258. The green box labels the extraction region of the soft, diffuse emission. The red dashed box marks the size of the zoomed inset in the right panel. {\it Right.} \chandra ACIS-S $0.5-7.0$ keV image of NGC 4258, zoomed on the nuclear region with the off-nuclear source clearly visible. Its spectrum has been extracted from the green circular region. The red dashed box marks the corresponding size of the zoomed inset in the left panel. In both panels, north is up and east to the left.}
   \label{fig:cxo}
\end{figure*}

NGC 4258 was observed multiple times by all major X-ray astronomical facilities in the last decades. Here we focus on the most recent \chandra ACIS-S, \xmm EPIC PN and MOS, \nustar FPMA and FPMB, and Swift/BAT observations. A summary of the observations considered, with their dates and cleaned exposure times, is shown in Table \ref{tab:info}. Together with this data, we will consider previously published results from the literature, regarding ASCA \citep{makishima94, reynolds00, terashima02}, BeppoSAX \citep{fiore01} and \suzaku \citep{reynolds09} observations. For all of the analysis we use the fitting package XSPEC \citep{arnaud96} v.12.11.1.

\subsection{\chandra}
NGC 4258 has been observed four times by \chandra with the ACIS-S instrument, for a total exposure time of 46 ks. The nucleus is so bright that its emission is piled-up even in a modest \chandra exposure \citep{youngwilson04}. Indeed, the majority of the ACIS-S exposure time was intended to study the anomalous arms, while very few ks of observations with short time frames ($0.4 s$ and $0.1 s$) could in principle be used to extract less-piled-up spectra of the nuclear source. Pile-up manifests as a spectral distortion, hardening the spectrum and making it unnaturally flat. Although there are {\it ad-hoc} models available in spectral fitting software to mitigate the effects of pile up, they rely on a few poorly known parameters \citep[see, e.g., the pile up model by][]{davis01}. Therefore, we have chosen not to increase the number of parameters in our modeling, and instead took advantage of the excellent angular resolution of \chandra to extract the spectra of the non-nuclear components that are very likely to contaminate the \xmm and \nustar data, due to their lower angular resolution and larger point spread function (PSF). \chandra observations were downloaded from the \chandra\xspace {\it Data Archive} and reprocessed with the \texttt{chandra\_repro} task in CIAO \citep{fruscione06} v.4.12. The \texttt{specextract} task was used to extract the spectrum of the circum-nuclear soft, diffuse emission and of an off-nuclear X-ray source, $2\farcs5$ south-east from the nucleus and already known in the literature \citep{wilson01, pietschread02, youngwilson04}.

\subsection{\xmm}
NGC 4258 was observed by \xmm multiple times, in particular for five epochs during the early 2000s \citep{fruscione05}, and once in 2006 \citep{reynolds09}. The \xmm observations were downloaded from the HEASARC archive and reduced with the SAS v.19.1.0. In particular, cleaned event files were created for both the PN and MOS cameras onboard XMM with the \texttt{epproc} and \texttt{emproc} tasks, respectively. No pileup was detected, as reported in \citet{fruscione05}. High background time intervals were selected using light curves in the $10-12$ keV energy range. The event files were thus filtered with these defined good time intervals, and adopting the standard FLAG=0 and pixel groups with pattern $\leq 4$ and $\leq 12$ for the PN and MOS cameras, respectively. Finally, source and background spectra, as well as ancillary files and responses, were extracted using the SAS task \texttt{xmmselect}. Source spectra were extracted from 15"-radius circular apertures, while background spectra were extracted from larger circles (with size ranging from 60" to 90") on the same detector chips. \texttt{epicspeccombine} was used to co-add MOS1 and MOS2 spectra for each of the six epochs, resulting in a total of 12 \xmm spectra. All the spectra have been rebinned to have at least 20 counts per bin with the HEASOFT tool \texttt{grppha}.

\subsection{\nustar}
\nustar observed NGC 4258 twice, for a total exposure time of $\sim 158$ ks. We downloaded the ObsIDs from the HEASARC archive, and reduced the observations with the NuSTARDAS package, using the standard \texttt{nupipeline} task to clean the raw data. The cleaned event files were used to extract the spectral products, through the task \texttt{nuproducts}. Source and background spectra were extracted from circles of 50" and 120" radius, respectively. The background regions were chosen on the same detector chip where the source spectra were extracted. All the spectra have been rebinned to have at least 20 counts per bin with the HEASOFT tool \texttt{grppha}.

\subsection{\bat}
We downloaded the \bat spectrum of NGC 4258 from the \bat 105-Month Hard X-ray Catalog \citep{ricci17BASS, oh18} \footnote{https://swift.gsfc.nasa.gov/results/bs105mon/609}, for a final broad band range $0.3-195$ keV.

\section{Off-nuclear contaminants}\label{sec:offnuc}
The deepest \chandra ACIS-S observation (ObsID 1618, see Table \ref{tab:info}) was used to extract the spectrum of the diffuse, extended soft X-ray emitting plasma in the nucleus of NGC 4258 and of the off-nuclear point source. 
While the former basically disappears at energies $E \gtrsim 2$ keV, contributing negligibly to the hard X-ray flux, the latter component is expected to contribute, to some extent, to the flux in the \xmm and \nustar energy bands.

\subsection{Soft, diffuse emission} \label{sec:soft}
As detailed in \citet{wilson01}, the clear diffuse X-ray emission coincident with NGC 4258 anomalous arms, historically detected also with radio and H$\alpha$ imaging, is due to galactic disk gas that has been shocked by mass motions driven by the out-of-plane radio jets. More recently, evidence for shocks has also been found in cold molecular gas \citep{ogle14}. Thus, much of the gas originally in the disk has been ejected into the galaxy halo in an X-ray hot outflow. As a consequence, the star formation in the central few kpc is rather low \citep[0.08 $M_\odot$/yr,][]{ogle14}. In addition, the anomalous arms themselves have been suggested to be free of star formation \citep{courtes93}. We extract the \chandra spectrum of the diffuse, soft emission along the anomalous arms of NGC 4258, from a rectangular region south-east of the nucleus (left panel of Figure \ref{fig:cxo}). The spectrum is grouped to have at least 20 counts per bin, and is fitted with a double \texttt{mekal} \citep{mewe85} component in XSPEC. The spectrum is equally well fit either by two different temperatures with solar abundance, or assuming only one component, in which case the abundance is sub-solar. To be consistent throughout the paper, we assume solar abundance and adopt two \texttt{mekal} components to describe this component \citep[e.g.,][]{reynolds09}.

\subsection{Off-nuclear point source} \label{sec:oa}
The presence of an off-nuclear point source, located 2\farcs5 to the south-east of the nucleus of NGC 4258, has been known for more than two decades \citep{wilson01, pietschread02, youngwilson04}. Despite this, its exact nature is still unknown. Its \chandra spectrum is extracted from a small circular region, carefully avoiding the emission from the active nucleus of NGC 4258 itself (right panel of Figure \ref{fig:cxo}). The spectrum is equally well fit by an absorbed power-law \citep[in which case we find results consistent with those of][]{youngwilson04}, and by a multicolor blackbody model (\texttt{diskbb} in XSPEC). In the first case, the source could be a background AGN with a fairly typical photon index ($\Gamma = 1.74^{+0.45}_{-0.41}$) obscured by a column density $N_{\rm H} = 2.9^{+2.5}_{-2.1} \times 10^{21}$ cm$^{-2}$ (which could be due to gas in NGC 4258 itself), while in the second case it could be an X-ray binary in the nuclear region of the galaxy, with a temperature $kT = 1.4^{+0.6}_{-0.3}$ keV, and whose flux is again absorbed by a column density in excess of $10^{21}$ cm$^{-2}$. The unknown nature of this source results in an uncertain extrapolation of its contribution to the X-ray flux at higher energies, depending on the spectral model assumed. To be conservative, we consider the absorbed power-law model (which provides the largest contribution at hard X-ray energy) fixing the best fit column density, photon index and power-law normalization to their best fit values. We have also verified that across the three \chandra observations in which the source is detected (spanning approximately one year), its $0.5-7$ keV flux is rather stable, showing possible variations at the $20\%$ level -- albeit consistent with being constant within the uncertainties. Its spectral shape is also constant within the uncertainties. However, the quality of the data in the other observations allowed only a rough assessment of the X-ray spectral shape of the off-nuclear point source to be done. Its observed $2-10$ keV flux is $F_{2-10} \sim 10^{-13}$ \fluxcgs, $\approx 2\%$ of the average observed flux in the same band from NGC 4258. 

\section{X-ray spectral analysis}\label{sec:analysis}

\begin{figure}
   \centering
   \includegraphics[width=0.5\textwidth]{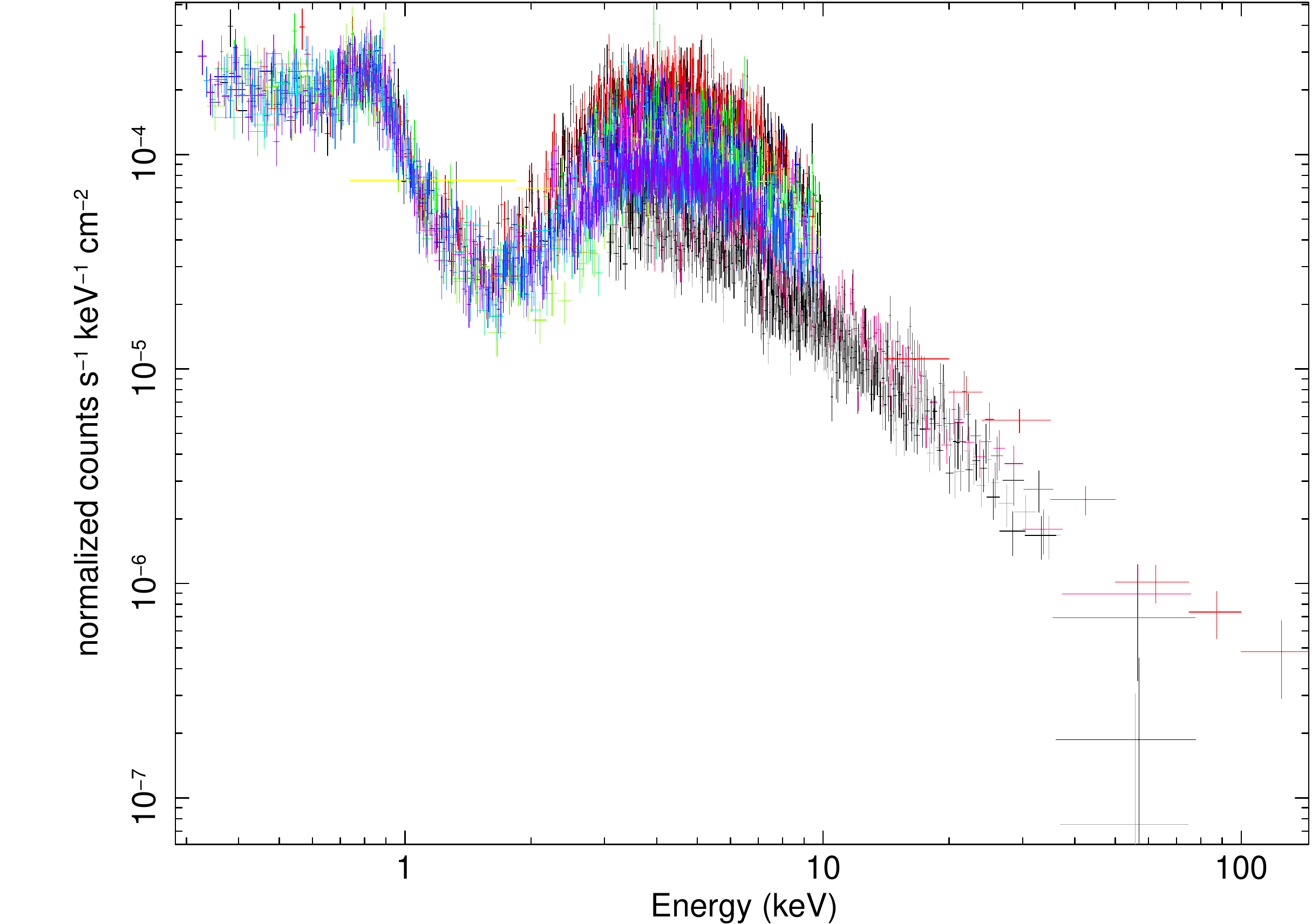}
   \caption{Broad band $0.3-195$ keV spectrum of NGC 4258, composed of the 17 spectra analyzed in this work (12 \xmm, 4 \nustar and 1 \bat) spanning more than 15 years of observations. While the soft emission $\lesssim 2$ keV is due to extended plasma and is constant throughout the years, the harder component shows significant variability across different observations. In particular, NGC 4258 was caught in its faintest state during the \nustar observations. The spectral turnover below $4-5$ keV is due to the moderate absorption, possibly due to the line of sight intercepting the warped maser disk.}
   \label{fig:spec}
\end{figure}

The broadband \xmm + \nustar + \bat spectra, covering almost three orders of magnitude in energy ($0.3-195$ keV) and more than 15 years in time, are shown in Figure \ref{fig:spec}. They show two peaks, around $\sim 4-5$ keV and $\sim1$ keV. While the latter peak is due to the diffuse plasma at large scales, the higher energy one is due to the absorbing column density affecting preferentially soft X-ray photons, hence producing the observed spectral curvature. Consistently with this interpretation, the soft ($\lesssim 2$ keV) emission is stable over the years, while the hard X-ray flux is variable by a factor of $\sim 10$ at $E > 3$ keV. Can this variability be accounted for by {\it flux variability} only, or is it more complex (i.e., a {\it spectral variability})? To answer this question, we will test two different setups for any given spectral model.

\subsection{Models setup}\label{sec:setup}

\begin{figure*}
   \centering
   \includegraphics[width=0.45\textwidth]{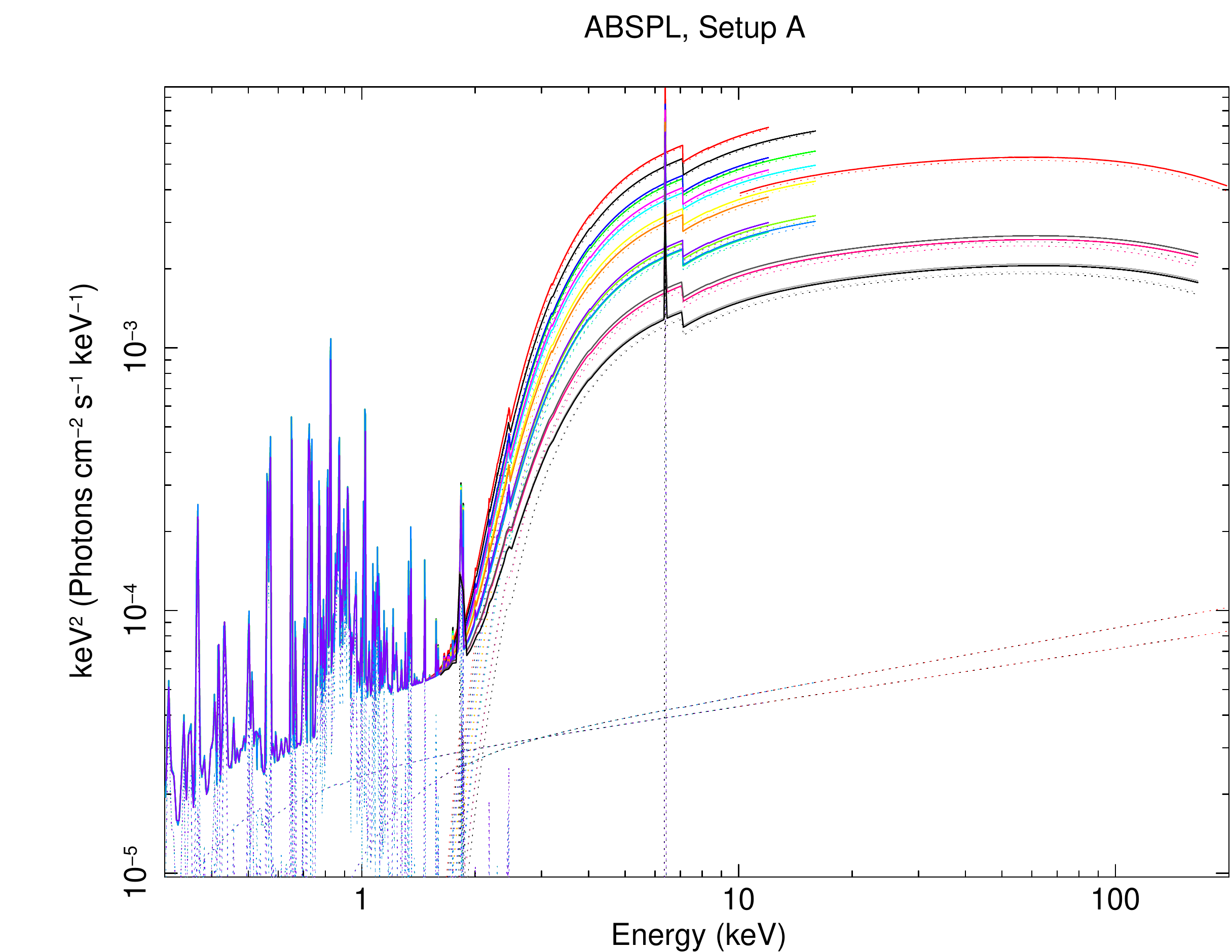}
   \includegraphics[width=0.45\textwidth]{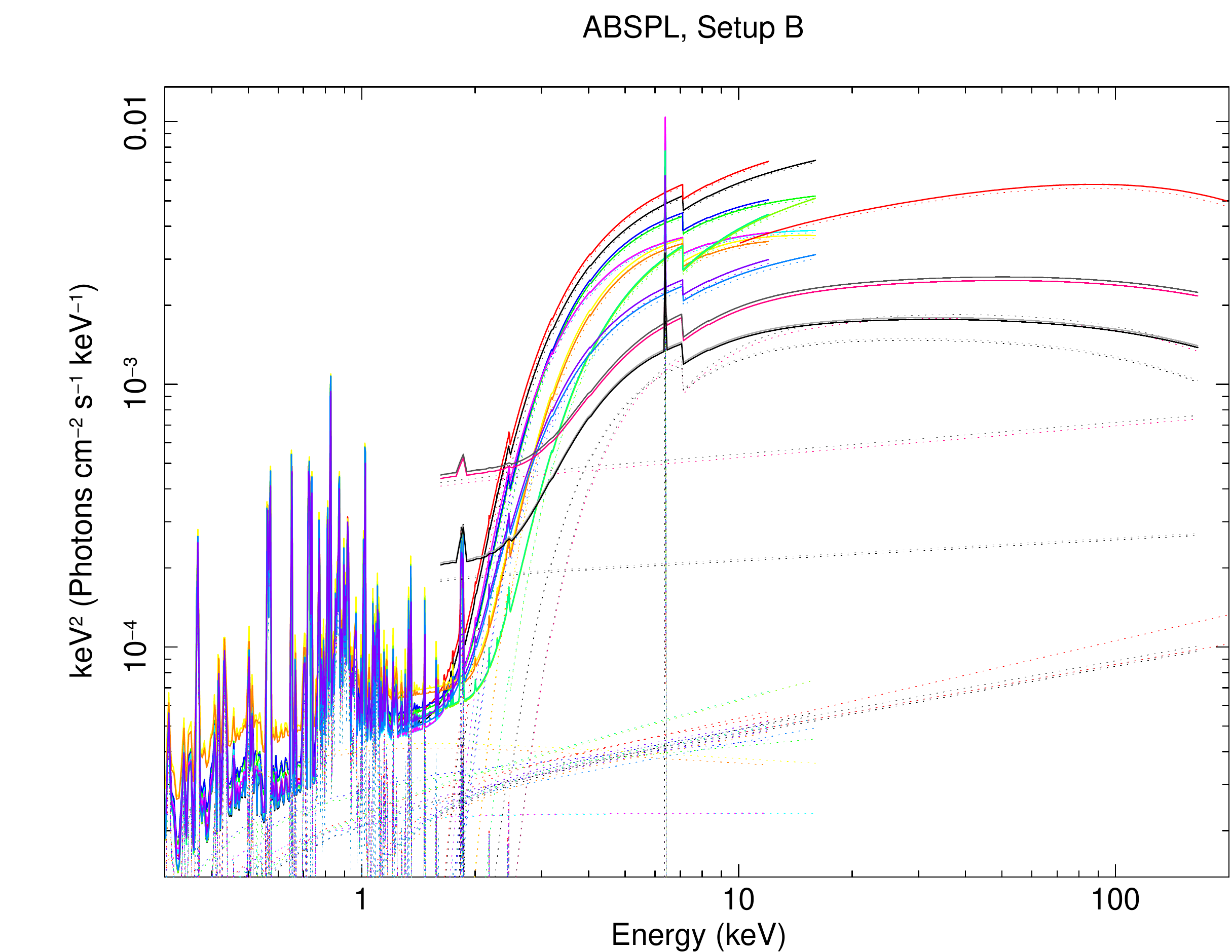}
   \includegraphics[width=0.45\textwidth]{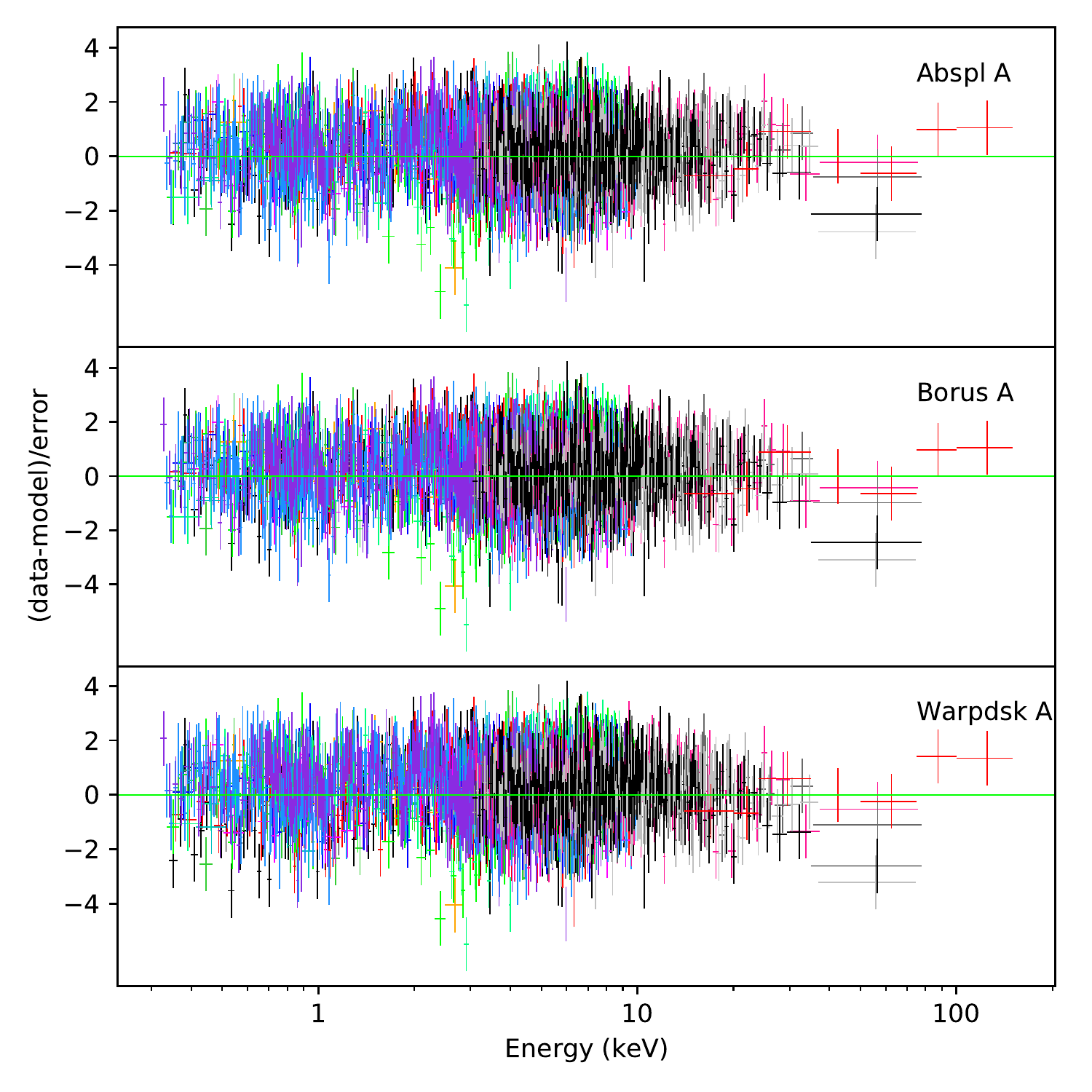}
   \includegraphics[width=0.45\textwidth]{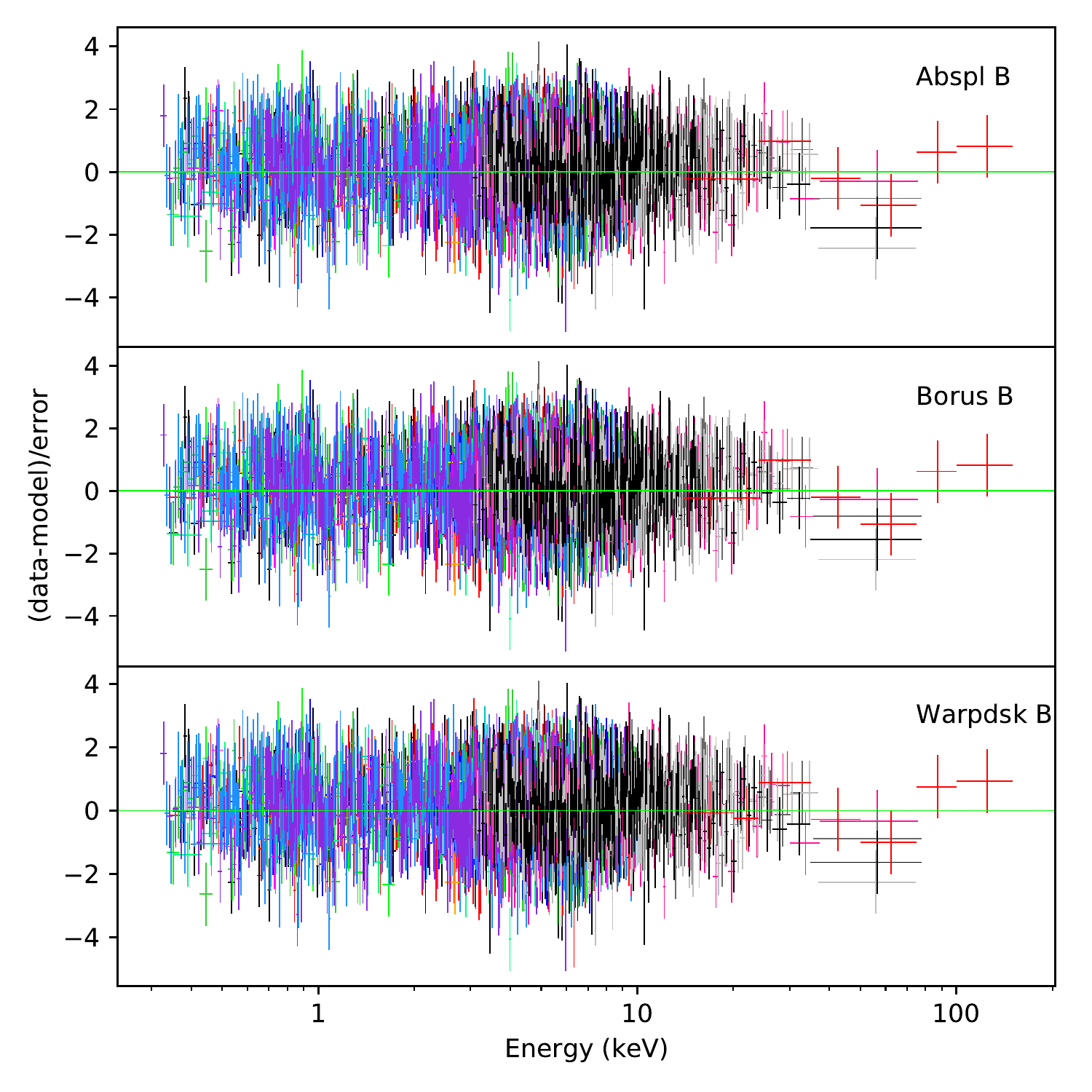}
   \caption{{\it Top row.} The two different setups assumed in the spectral analysis are shown here, adopting the ABSPL model as an example. The left panel refers to setup A, in which the spectral parameters are the same for all the data sets (labeled with different colors), modulo a normalization constant that encapsulates the nuclear flux variability of the hard X-ray emission alone. Setup B is shown in the right panel, for which the diversity in spectral slope, column density and normalization can be appreciated.
   {\it Bottom row.} Data vs model residuals, as fit with the setup A (left panel), and B (right panel): from top to bottom: ABSPL, BORUS and WARPDSK models, respectively.}
   \label{fig:setups}
\end{figure*}

In the first setup, dubbed setup A, we assume that the observed variability is due only to flux variability; thus, we fit all the 17 spectra using the same set of parameters, allowing for a free normalization constant for the nuclear component alone, relative to the first epoch, December 2000 (see the left panel of Figure \ref{fig:setups})\footnote{When simultaneously fitting datasets from different instruments, a multiplicative cross-calibration constant is usually considered. However, in this case we do not include a cross-calibration constant in Setup A, since the cross-calibration uncertainty is much smaller \citep[$\sim10-15\%$;][]{madsen15,madsen17} than the observed flux spread, and the fit would be insensitive to its value.}. In the second setup, dubbed setup B, we instead allow for spectral variability, leaving the column density, photon index and normalization of the coronal power-law free for each epoch\footnote{In setup B, we have fitted at each epoch the cross-calibration constant among cameras of the same instrument (see Table \ref{tab:B}).} (right panel of Figure \ref{fig:setups}). In both setups, the soft emission is modeled in the same way. In all the analysis, the Galactic absorption is fixed to $N_{\rm H,Gal} = 4.19\times 10^{20}$ cm$^{-2}$ \citep{hi4pi16}.

The data requires multiple components to reproduce the broadband spectrum of NGC 4258. First, a soft X-ray component to fit the emission below 2 keV. This has been extensively studied, and is usually fitted with two plasma emission models, as discussed in Section \S \ref{sec:soft}. Then, the nuclear hard X-ray emission is generally well described, to first order, by the ``typical'' spectrum of a local AGN, i.e. a power-law modified at low energies by some intervening absorption along the line of sight, and declining at high energies as an approximately exponential cutoff. We fix the high energy cutoff at the median observed value for local Seyfert IIs, i.e. $290$ keV, which is found to be consistent with that of Seyfert Is \citep{balokovic20}. When some degree of absorption is present, both a scattered power-law at soft energies, mirroring the primary one, and a narrow fluorescent Fe K$\alpha$ line at $E\approx 6.4$ keV are also expected. In this particular case, the controversial significance of the line detection which has been somewhat debated in the literature \citep[see, e.g.,][]{reynolds09}, further motivated us to include in our baseline model a Gaussian line component. Finally, an absorbed power-law describing the contamination from the off nuclear point source is taken into account and it is fixed by the best fit values derived from the \chandra spectrum (\S \ref{sec:oa}). In XSPEC notation, this baseline model (dubbed ``ABSPL") reads out as 

\begin{multline}
    \overbrace{\texttt{TBabs}}^{\text{Galactic $N_{\rm H}$}}\times \{ 
    \overbrace{\texttt{zphabs}\times \texttt{cabs}\times \texttt{cutoffpl} + \texttt{zgauss}}^{\text{Intrinsic absorbed emission and Fe K$\alpha$}} + \\ 
    + \underbrace{\texttt{mekal} + \texttt{mekal} + \texttt{zpowerlw}}_{\text{Soft plasma and scattered power-law}} + \underbrace{\texttt{phabs}\times \texttt{powerlw}}_{\text{Point source contaminant}} \}.
\end{multline}

The ABSPL model provides a very good phenomenological description of the data. However, more recent self-consistent models have been developed, based on Monte Carlo simulations of the radiative transfer of X-ray photons emitted by a central source and propagating through some neutral medium with a given assumed geometry \citep{ikeda09, murphy09, brightmannandra11, balokovic18}. Therefore, we will also fit the broad band spectrum with one such more physically motivated descriptions, the \texttt{Borus02} (BORUS) model of \citet{balokovic18}. The assumed geometry of the obscuring medium is a homogeneous sphere with polar cutouts, in which opening angle $\theta_{oa}$ of the torus, measured from the polar axis of the system, is an adjustable parameter. The covering factor CF, defined as the fraction of the sky as seen from the corona obscured by the torus, in this model is $\text{CF}= \cos{\theta_{oa}}$, and allows to explore both sphere-like and disk-like geometries. In XSPEC, its implementation is similar to the ABSPL model:

\begin{multline}
    \overbrace{\texttt{TBabs}}^{\text{Galactic $N_{\rm H}$}}\times \{ 
    \overbrace{\texttt{zphabs}\times \texttt{cabs}\times \texttt{cutoffpl}}^{\text{Intrinsic absorbed emission}} +\hspace{-0.2cm}\overbrace{\texttt{Borus02}}^{\text{Torus reprocesssing}}+\\ 
    + \underbrace{\texttt{mekal} + \texttt{mekal} + \texttt{zpowerlw}}_{\text{Soft plasma and scattered power-law}} + \underbrace{\texttt{phabs}\times \texttt{powerlw}}_{\text{Point source contaminant}} \}.
\end{multline}

The warped megamaser disk of NGC 4258 has been robustly studied and mapped in detail \citep[e.g.,][]{herrnstein98, herrnstein05, humphreys13} and it has been suggested to be responsible for the obscuration along the line of sight \citep{fruscione05}. \citet{buchner21} recently 
proposed a spectral model, based on a warped disk geometry, which we will refer to as WARPDSK. The total emission of WARPDSK is the sum of two components: the primary one (\texttt{warpeddisk.fits}) consisting of photons coming directly from the central source and accounting for Compton scattering and absorption from the warped disk, and a second one (\texttt{warpeddisk-omni.fits}) being the so-called omni-directional scattered component due to photons randomly scattered into our line of sight. This warm mirror emission is the angle-averaged (omni-directional) spectrum, predominantly comprising the incident power-law from unobscured lines of sight, and effectively replaces the scattered power-law of the ABSPL and BORUS models. The free parameter \texttt{diskfrac} basically represents the ``strength'' of the warp: a value \texttt{diskfrac} close to unity suggests a strong warp,  while a smaller value hints to a flatter geometry. As such, it can be interpreted as a CF as well, although it is not trivial to directly compare the CF from BORUS with \texttt{diskfrac} in WARPDSK. All the parameters of the omni-directional component are linked to those of the main one -- but its normalization, which effectively sets the scattered fraction.

In XSPEC notation, the WARPDSK model is implemented as follows:

\begin{multline}
    \overbrace{\texttt{TBabs}}^{\text{Galactic $N_{\rm H}$}}\times \{ 
    \overbrace{\texttt{zphabs}\times \texttt{cabs}\times \texttt{warpeddisk}}^{\text{Intrinsic absorbed emission}} + \\ +\overbrace{\texttt{warpeddisk-omni}}^{\text{emission scattered into the LOS}}
    + \overbrace{\texttt{mekal} + \texttt{mekal}}^{\text{Soft plasma emission}} + \\ 
    +\underbrace{\texttt{phabs}\times \texttt{powerlw}}_{\text{Point source contaminant}} \}.
\end{multline}

\subsection{Results}\label{sec:results}

The simplest model is the ABSPL with the same parameters for all 17 spectra, modulo a free constant to account for flux variability only (setup A). The reduced $\chi^2$ is already acceptable ($\chi^2/\text{dof} = 3976/3761$), with the two \texttt{mekal} components, with significantly different temperatures, describing sufficiently well the soft emission $< 2$ keV (see \S\ref{sec:soft}). The hard X-ray power-law is obscured by a column of almost $10^{23}$ cm$^{-2}$ and has a slope consistent with a typical value $\Gamma \sim 1.8$.

As expected, the constants encapsulating the flux variability are significantly different from unity (set by the flux of the first \xmm  observation) and interestingly, show a decreasing trend with time: the flux at the last \nustar epoch is around $\sim 25\%$, while the average \bat flux is roughly 65\%\footnote{Taking the \bat flux as a reference, this means that the first \xmm and last \nustar epochs caught NGC 4258 with a flux $\sim 54\%$ higher and $\sim 62\%$ lower than the average, respectively.}. The other two models, BORUS and WARPDSK, give results consistent with those of the ABSPL one ($\chi^2/\text{dof} = 3969/3760$ and $\chi^2/\text{dof} = 3976/3760$, respectively; bottom left panel of Figure \ref{fig:setups}), both in terms of the soft X-ray emission and the hard X-ray coronal slope and obscuration. Across all models, the scattered power law component needed to fit the spectrum in the $1-3$ keV range gets washed out at lower energies, due to the thermal plasma emission. The only noticeable difference, although within the uncertainties, is in the intrinsic normalization, which results $\sim 50\%$ higher adopting the WARPDSK model. Despite this, the combination of the higher normalization with the softer photon index of the WARPDSK model (see \S \ref{sec:longterm}) gives fully consistent intrinsic $2-10$ keV flux (and therefore luminosity) among models. A detailed comparison with results from previous work \citep[e.g.,][]{fruscione05, reynolds09, kawamuro16, panagiotouwalter19, osorioclavijo21} is not trivial, especially due to different, more recent models being adopted here. In general, however, our spectral parameters are consistent with those found in the literature.

The ABSPL model allows to make an assessment over the significance of the Fe K$\alpha$ emission line, whose rest frame energy has been fixed to 6.4 keV: its average equivalent width (EW) results rather low (EW $= 45\pm12$ eV), similar to the EW of the Si K$\alpha$ line identified at $\sim 1.84$ keV. If the energy of the Fe line feature is left free to vary, the $\chi^2$ is not significantly improved and the energy is well constrained ($E_{\rm FeK\alpha} = 6.40\pm0.03$) around the expected value for neutral Fe K$\alpha$ emission. 
One clear advantage of both the BORUS and WARPDSK models is their self-consistency in treating radiative transfer within the assumed geometry. As previously mentioned, BORUS has the CF of the torus as a free parameter: when fitting for the CF, we find an upper limit of $\text{CF}<0.16$, strongly suggesting that the maser disk is responsible for the obscuration and supporting previous independent claims \citep[e.g.,][]{fruscione05}. Furthermore, if we leave both the \nh (representing the line of sight column density) and the parameter $\log{N_{\rm H,Tor}}$ of BORUS \citep[interpreted as a global average \nh; see][]{balokovic18} free, slightly different values are returned, suggesting the obscuring medium to be clumpy. The fit with the WARPDSK model in setup A returns a \texttt{diskfrac} parameter $\sim 0.9$, suggesting a strong warp. The detailed results for setup A are summarized in Table \ref{tab:A}. 

The most interesting results are obtained when refining the analysis by leaving the photon index, column density and coronal flux normalization free for each epoch, i.e. adopting setup B. This setup allows also to explore if and how the spectral shape and X-ray luminosity of the source have changed over the years. For simplicity, when fitting with BORUS we linked the global average \nh of the torus and the line of sight \nh. In general, the three models return acceptable fits with much better reduced $\chi^2$ with respect to the corresponding setup A. This is not merely due to the increase of the free parameters, as the $\Delta\chi^2/\Delta\text{dof} = 356/28, 345/21, 323/23$ for the ABSPL, BORUS and WARPDSK models, respectively (bottom right panel of Figure \ref{fig:setups}). This supports evidence for spectral variability between different epochs.
Indeed, clear column density and spectral slope variations are detected, independently of the spectral model assumed, and will be further discussed in Section \S\ref{sec:variability}. At any given epoch, the three models give consistent results as shown in Figure \ref{fig:var} and Table \ref{tab:B}.
Both the CF parameter of BORUS and the \texttt{diskfrac} parameter of the WARPDSK model were free to vary but kept linked for all epochs, and the fitted values are significantly different from those of setup A, suggesting a slightly larger CF (CF$=0.37\pm0.14$), and a much less prominent warping of the disk ($\texttt{diskfrac}=0.25\pm0.01$), respectively. This last result could be somewhat expected. First, the larger number of free parameters in setup B allows to better fit the spectral shape at each epoch, thus returning a new value for the warp strength which is significantly different from before. Furthermore, as detailed in \citet{buchner21}, WARPDSK is tailored for heavily Compton-thick AGN, as the vast majority of disk megamasers turn out to be. Hence the megamaser disk, by construction, has an assigned column $\log{N_{\rm H}/\text{cm}^{-2}} = 25$ to provide the extreme obscuration often observed in the best studied local megamasers \citep[e.g.,][]{arevalo14,puccetti14,bauer15,masini16}. In this particular case, however, in order to account for the relatively low column density affecting the spectrum, WARPDSK is forced by the fixed inclination of the inner disk \citep[72\degree,][]{humphreys13} to graze the warp, fitting the \texttt{diskfrac} parameter to the lower reported value.

\subsection{NGC 4258 in the context of local, low-$\lambda_{\rm Edd}$ AGN}\label{sec:context}

Disk megamasers, when compared with the local \bat-selected AGN population, generally occupy the high-end of the $\lambda_{\rm Edd}$ distribution \citep[see, e.g., Figure 8 of][]{masini19}. In this regard, NGC 4258 is definitely peculiar with respect to other disk megamasers, although other \bat-selected AGN show accretion rates comparable with that of NGC 4258. Thus, it is interesting to consider this source in a broader context, comparing its properties with those of other local, low-Eddington ratio AGN. The sample of 81 local AGN observed by \nustar with $\log{\lambda_{\rm Edd} < -3}$ presented by \citet{osorioclavijo21}, in which NGC 4258 is included as a LINER, is the ideal sample to make such a comparison. In terms of black hole mass, bolometric luminosity and Eddington ratio, NGC 4258 is fully consistent with the average values found for their sample of local LINERs. The same holds true also regarding its X-ray properties: both the photon index and the average column density are consistent with their respective averages of the LINER sample. In general, the results about NGC 4258 reported in this work and \citet{osorioclavijo21} are consistent, despite applying different models, and we also do not find any significant contribution of a reflection component in the hard X-ray spectrum of NGC 4258. These findings support the hypothesis that at low accretion rate the standard picture of the torus may break down. In the specific case of NGC 4258, the missing reflection component may be a consequence of the low covering factor of the warped maser disk surrounding the AGN.

\section{Short and long term variability}\label{sec:variability}

\begin{figure}
   \centering
   \includegraphics[width=0.5\textwidth]{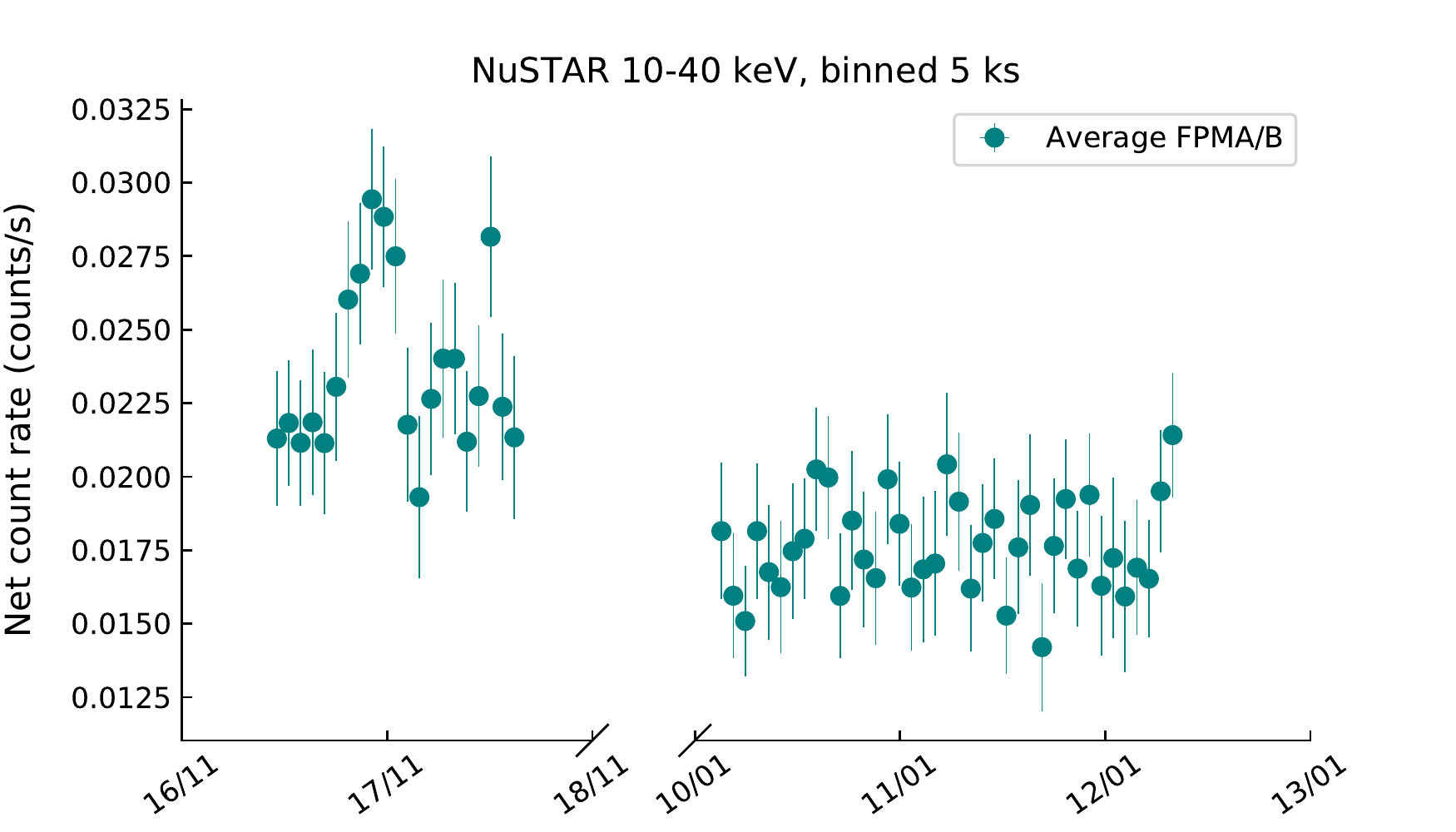}
   \caption{\nustar (average of FPMA and FPMB) $10-40$ keV light-curve across two observations, showing the background-subtracted count rate in bins of 5 ks. The shorter observation of November 2015 displays variability on short timescales (a few ks), which are intrinsic to the source (not due to obscuration).}
   \label{fig:lc}
\end{figure}

As shown in the previous Section, setup B allows to keep track of the spectral parameters and the luminosity of NGC 4258 as a function of time. Moreover, the last two \nustar epochs allow to explore also the intrinsic, hard X-ray emission from the putative corona. In the following, we thus focus on both the short and long term variability of NGC 4258.

\subsection{Short term variability}

We take advantage of the \nustar data to explore the intrinsic hard X-ray variability (i.e., not caused by variable absorption) of the AGN. The background-subtracted $10-40$ keV \nustar light curve for both ObsIDs is shown in Figure \ref{fig:lc}, where data of the two focal plane modules have been averaged together and binned with a 5 ks bin size. First, we note again that the average, net count rate from NGC 4258 has decreased between the two \nustar observations by roughly $\sim20\%$. While the most recent data do not show any particular trend, a clear peak is visible in the first dataset around 17-Nov-2015. We have extracted the \nustar spectrum of the ``mini-flare'' and compared it with the quiescent spectrum, finding no significant difference in either spectral slope or column density. This could be due to the relatively weak intensity of the peak, which is actually less than a $50\%$ increase in count rate.

Figure \ref{fig:lc} shows that a rise time of $\sim 20$ ks is observed for the mini-flare on 17-Nov-2015. This timescale looks halfway between those found by \citet{reynolds09}, who detected both a significant brightening of NGC 4258 during the 2007 {\it \suzaku} exposure with a timescale of $\sim50$ ks, and lower amplitude fluctuations over much shorter timescale of $\sim5$ ks. Assuming that variability is associated with accretion disk processes, \citet{reynolds09} equated these variability timescales with the dynamical time of a Keplerian disk $\Omega^{-1} = \sqrt{r^3/GM}$, to put constraints over the size of the X-ray emitting region. Under the same assumption, the observed timescale of $\sim 20$ ks corresponds to an emitting region size of $\sim 20r_g$, where $r_g = GM/c^2$ is the usual definition of the gravitational radius.

\subsection{Long term variability and Fe K$\alpha$ line}\label{sec:longterm}

\begin{figure*}
   \centering
   \includegraphics[width=\textwidth]{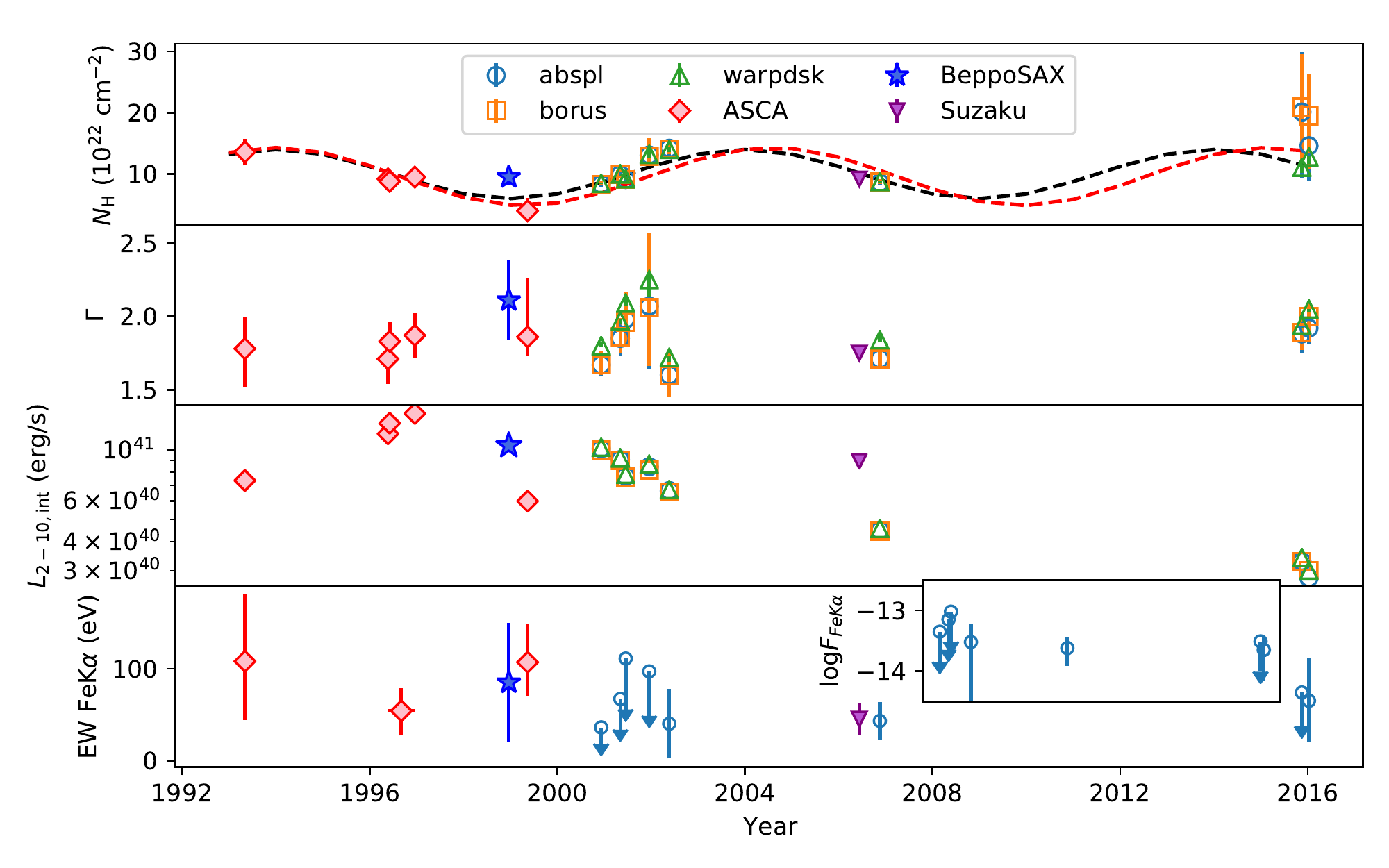}
   \caption{Long term evolution of the main spectral parameters. From top to bottom: column density in units of $10^{22}$ cm$^{-2}$, photon index $\Gamma$ (generally  between $1.6-2.0$), intrinsic (de-absorbed) $2-10$ keV luminosity, EW of the Fe K$\alpha$ line (only for the ABSPL model and for values from the literature; the inset shows the logarithmic flux of the Fe K$\alpha$ line in the ABSPL model, for the data considered in this work). The values referring to the three models employed in our analysis are labeled with different empty markers and colors (blue circles, orange squares and green upward triangles for the ABSPL, BORUS and WARPDSK models, respectively), and are fully consistent with each other. Red diamonds, the blue star and the purple downward triangle refer to ASCA \citep{reynolds00, terashima02}, BeppoSAX \citep{fiore01} and \suzaku \citep{reynolds09} observations, respectively. In the top panel, the dashed black line is a simple sinusoid with a period of 10 years, and is not a fit to the data. Most notably, the intrinsic de-absorbed 2-10 keV luminosity shows an approximately steady decrease of more than a factor of three over $\sim20$ years.}
   \label{fig:var}
\end{figure*}

The spectrum of NGC 4258 shows moderate flux variability among the \xmm and \nustar observations, as previously reported in the literature, and as demonstrated also by our analysis. Therefore, we collected the spectral parameters of the three models employed in Section \S\ref{sec:analysis} as a function of time, along with previous results from the literature. This allows us to explore the long term behavior of NGC 4258, spanning 23 years, between 1993 and 2016.

The first panel from the top of Figure \ref{fig:var} shows the variability of the the X-ray obscuring column density \nh. The column density is always around $\sim 10^{23}$ cm$^{-2}$, with fluctuations of a factor of $\sim2$. As detailed in \citet{fruscione05}, \nh has been observed to increase over a few months timescale in 2001, with an associated length scale of about $10^{15}$ cm, very similar to the expected scale height of a standard thin accretion disk. We find a fully consistent trend in the first five \xmm epochs analyzed here, in which \nh increases from $\sim 8\times 10^{22}$ cm$^{-2}$ to $\sim 14\times 10^{22}$ cm$^{-2}$ over $\Delta t = 530$ days. Assuming that the absorbing gas is located at the radius where the warp is expected to cross the line of sight, i.e. $R \sim 0.3$ pc, and that it orbits in a Keplerian rotation the central mass of $M = 4 \times 10^{7}$ M$_\odot$, the orbital velocity is $v = \sqrt{GM/R} \sim 760$ km s$^{-1}$. The characteristic size is then $s = v\Delta t \sim 3.5 \times 10^{15}$ cm. Assuming the typical density for masers to occur $\rho \sim 10^8$ cm$^{-3}$, and the column density variation $\Delta N_{\rm H} \sim 6 \times 10^{22}$ cm$^{-2}$, the associated depth variation (i.e., size along the line of sight) is $\Delta L \sim \Delta N_{\rm H}/\rho \sim 6 \times 10^{15}$ cm, which is roughly twice the previously derived characteristic size $s$. This increase in column density could be obtained either by doubling the number of clouds along the line of sight, or doubling the density of the absorbing medium. Interestingly, the column density seems to oscillate more or less regularly on a timescale of about 10 years, as suggested by the sinusoidal function with a period of 10 years we have plotted over the data. Although the reduced $\chi^2$ of fitting a simple sinusoidal function to the data is large enough to reject the null hypothesis ($\chi^2_{\nu} \sim 2.5$), a simple linear fit with roughly constant column density returns a much worse result ($\chi^2_{\nu} \sim 6.8$). If this remarkable, oscillatory trend is confirmed, its period ($\sim 10$ years) is much shorter than the orbital period at the warp radius ($\sim 2600$ years), and it may suggest periodic density variations in the absorbing gas, happening on a temporal scale of 5-10 years. This timescale corresponds to a linear scale of $1-2 \times 10^{16}$ cm ($0.003-0.006$ pc) at the warp radius. Interestingly, spiral density waves have been proposed to explain the periodicity and clustering of the masers in NGC 4258 \citep{humphreys08}. In the same paper, \citet{humphreys08} show that massive He I stars with a mass of $50-100 M_\odot$, comparable to those found in the Galactic center \citep{genzel03}, would be able to create gaps in the maser disk, with a size consistent with that observed here, and potentially responsible for the column density fluctuations.
The second panel from the top shows the long term evolution of the photon index $\Gamma$. It is generally measured between $1.6-2.0$, and there is no significant variability. We note that, at any given epoch, the WARPDSK model tends to return systematically softer (although statistically consistent) photon indices with respect to the ABSPL and BORUS models. The third panel of Figure \ref{fig:var} shows the intrinsic, absorption-corrected $2-10$ keV luminosity: after the early 2000s, the luminosity has been steadily decreasing, by a factor of $\sim 3$. The 2007 {\it \suzaku} observation seems to be the exception to this trend but, as mentioned earlier, \citet{reynolds09} caught NGC 4258 in a bright flux state during this observation. Since the luminosity is already corrected for variable absorption, the data strongly suggest that NGC 4258 is constantly getting fainter during the period considered.
It is interesting to compare the intrinsic $2-10$ keV luminosity derived here from that expected from the well known $12 \mu$m -- $2-10$ keV correlation. NGC 4258 has a nuclear $12 \mu$m luminosity $\log{L_{12 \mu m}/\text{erg s}^{-1}} = 41.26 \pm 0.05$, from which an expected $\log{L_{2-10 \text{keV}}/\text{erg s}^{-1}} = 41.03\pm0.04$ is derived \citep{asmus15}. This luminosity is consistent with that measured until the early 2000s, while the MIR observations described in \citet{asmus15} refer to the years 2010-2011, when NGC 4258 was already a factor of $\sim$ two fainter, according to our results. If there are no sources of contamination contributing significantly to the 12 $\mu$m flux, and if NGC 4258 has kept steadily decreasing its X-ray luminosity, this result could suggest a lag of the MIR with respect to the X-ray emission of at least a decade, thereby placing the MIR-emitting dust on a scale of $\sim 3$ pc.

Finally, the behavior of the Fe K$\alpha$ line (bottom panel of Figure \ref{fig:var}) is not easy to interpret. While the first ASCA observations suggested the presence of a moderately prominent emission line with an EW around $\sim 100$ eV, the subsequent \xmm observations we have considered in this work set only upper limits and never firmly detect the line, apart from the 2007 epoch \citep{reynolds09}, in which the line has in any case a modest EW. In the two most recent \nustar epochs, the line has been marginally detected only in the last (and longer) observation. This could be due to the more stable and systematically lower count rate of the source during the last \nustar epoch, as shown in Figure \ref{fig:lc}, which would make the line stand out more easily. If this hypothesis is correct, this would also imply that the Fe line is produced on large scales ($\sim$ pc scale) and lags significantly behind the continuum, differently from the inner disk origin interpretation of \citet{reynolds09}. Independently of its origin interpretation, the line EW from the \nustar epochs is consistent with that reported by \citet{reynolds09}. The inset in the last panel of Figure \ref{fig:var} shows the logarithmic flux of the Fe K$\alpha$ line for the spectra that we have analyzed in this work, derived through the XSPEC command \texttt{cflux}. The behavior of the line flux is fully consistent with that seen for its EW.

\section{Discussion}\label{sec:discussion}

The comparison of the X-ray spectra of NGC 4258 across two decades has shown that its X-ray properties have changed through the years both due to factor of $\sim$ two variations of the absorbing column density, plausibly associated with the dusty megamaser disk, and to intrinsic changes in the emission from the central engine -- most notably the coronal luminosity, which appears to have steadily decreased by a factor of $\sim$ three across $\sim15$ years, the only exception being the \suzaku observation which caught the source in a high state \citep{reynolds09}.

The short variability timescale ($\sim 20$ ks) observed in the hard X-ray \nustar data suggests that the variations might be due to changes in the accretion rate, in turn related to the rate of energy deposition in the corona. Accretion rate variability would also explain the long term decrease in intrinsic luminosity observed through 15 years of observations. The viscous timescale of a standard thin disk is generally much longer than few years. On the other hand, the SEDs of low luminosity AGN such as NGC 4258 are often explained in the context of RIAFs, i.e. with a combination of truncated \citet{shakurasunyaev73} disks with inner ADAFs and relativistic jets. Past efforts have successfully explained the broadband SED of NGC 4258 with such a model, although with their own shortcomings \citep[e.g.,][]{yuan02, wu13}. Interestingly, the viscous timescale is $\propto (r/H)^2$, thus becoming significantly shorter for an ADAF \citep{narayanyi95}. Both the \citet{shakurasunyaev73} and the RIAF models are in turn part of the so-called standard and normal evolution (SANE) models, in which the magnetic field is assumed to not significantly impact the dynamics of the disk. Opposed to SANE there are magnetically arrested disk models \citep[MAD;][]{narayan03}, expected to form when the accretion flow is supplied with a sufficient amount of magnetic flux. Unfortunately, it is difficult to observationally disentangle the two families of models \citep{xiezdziarski19}.

Furthermore, given the above results is not trivial to even discriminate between a RIAF and a standard thin disk. Indeed, the absorption-corrected X-ray luminosity $L_{\rm X}$, combined with archival determinations of the bolometric luminosity, implies a bolometric correction of $k_{\rm bol} \sim 20$, intriguingly typical of Seyferts powered by accretion through thin disks \citep{lusso12, duras20}. We note, however, that NGC 4258 could nonetheless be in a ``sweet spot'' in which the bolometric correction is consistent with that of a typical Seyfert, being it still powered by a RIAF \citep{nemmen14}. Moreover, its average photon index $\Gamma \sim 1.8$ is consistent with the typical value of the broader AGN population \citep[e.g.,][]{ricci17BASS}. Thus, our work suggests that NGC 4258 is a standard, low luminosity Seyfert II, despite its accretion rate in Eddington units being well within the expected RIAF regime.

\begin{figure}
   \centering
   \includegraphics[width=0.5\textwidth]{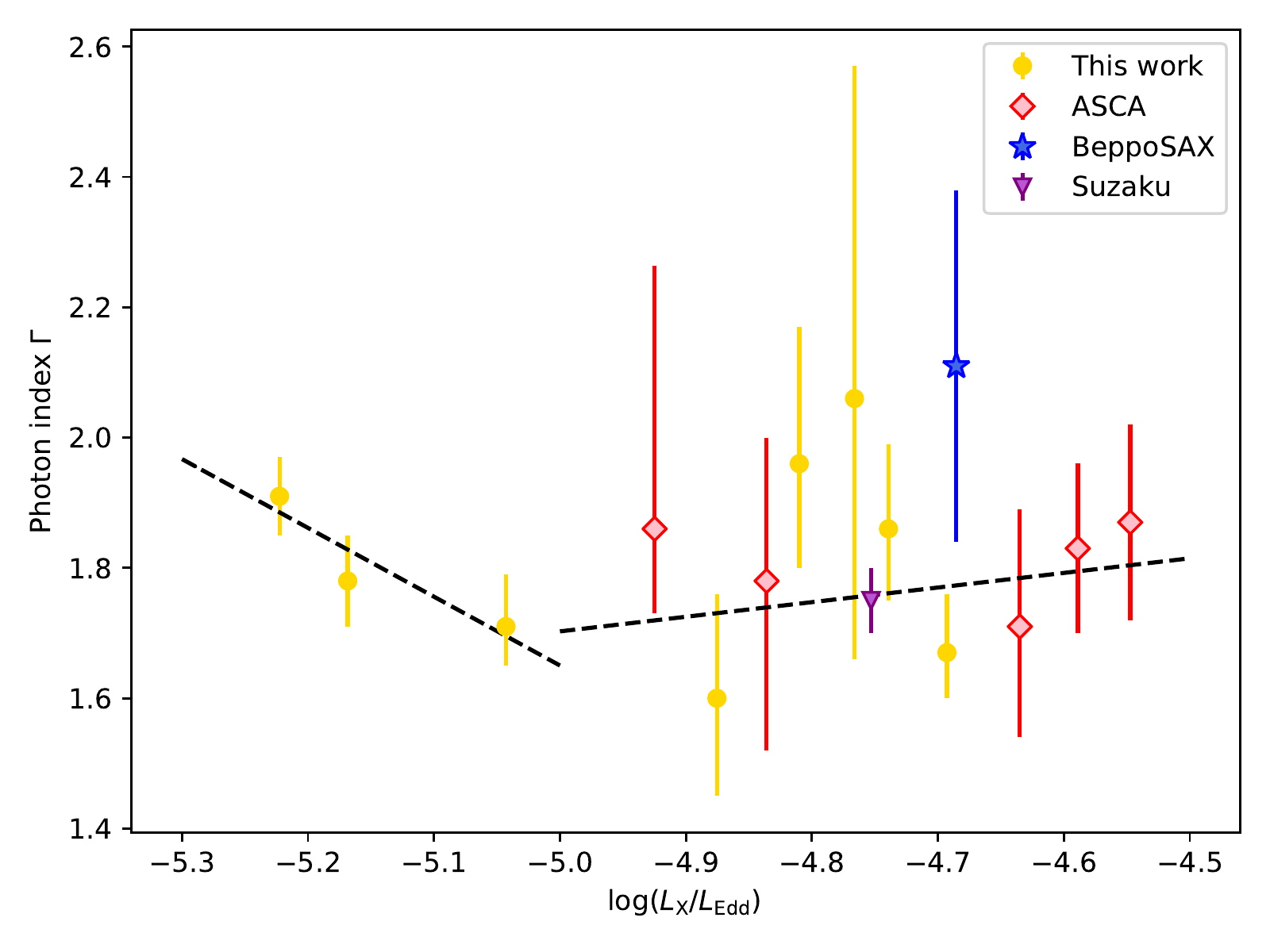}
   \caption{Photon index $\Gamma$ as a function of the X-ray Eddington ratio $\lambda_{\rm X}$ ($\lambda_{\rm X} = L_{\rm X}/L_{\rm Edd}$). Yellow points, red diamonds, the blue star and the purple triangle refer to the data analyzed in this work with the BORUS model, and to archival ASCA, BeppoSAX and \suzaku data, respectively. The black dashed lines are simple linear fits below and above the separation value of $\log{\lambda_{\rm X}} = -5$ (broadly corresponding to a canonical Eddington ratio of $\approx 10^{-4}$ for a bolometric correction of order $10$), showing two possible regimes.}
   \label{fig:gammaEddrat}
\end{figure}

Further discussion arises considering the X-ray photon index $\Gamma$ as a function of X-ray Eddington ratio $\lambda_{\rm X} = L_{\rm X}/L_{\rm Edd}$ (Figure \ref{fig:gammaEddrat}). Previous work has found somewhat conflicting results related to the $\Gamma-\lambda_{\rm Edd}$ correlation: \citet{brightman13} and \citet{brightman16} found a positive correlation between these two quantities, both for unobscured and heavily obscured AGN. On the other hand, \citet{trakhtenbrot17} found a much shallower correlation using a large, local \bat-selected sample. Analyzing a sample of low-luminosity AGN, \citep{gucao09} found instead an anti-correlation, similar to what is observed for X-ray binaries in the hard state \citep[e.g.,][]{liu19}. In our case, the spanned range in $\lambda_{\rm X}$ is around one order of magnitude, and there seems not to be any significant correlation between the two quantities considering the whole range. A possible anti-correlation may exist for $\log{\lambda_X} < -5$, similar to what is observed in other low-luminosity AGN \citep[e.g.,][]{kawamuro16}, and may suggest a transition between hot and cold accretion flows below and above a critical value $\log{\lambda_X} = -5$. Assuming a typical bolometric correction of order $\sim20$, this critical value would roughly correspond to $\lambda_{\rm Edd} \sim 10^{-4}$, which appears to be somewhat low for the transition between cold and hot accretion flows. However, this may suggest a MAD scenario to be in place, given the lower accretion rate at which a cold solution can still exist, with respect to SANE models \citep{xiezdziarski19}. In summary, if the change of slope in the $\Gamma-\lambda_{\rm X}$ plane is real, we may be witnessing the switch between cold and hot accretion flows in NGC 4258. Of course, at $\log{\lambda_{\rm X}} \lesssim -5$ there are just three data points at the moment, so future observations are highly warranted to support or discard this finding, especially if the trend of steady decrease in X-ray luminosity will be confirmed.

\section{Conclusions}\label{sec:conclusions} 

In this paper, we have analyzed the accretion properties of a well studied, nearby low-luminosity AGN, NGC 4258.
We have collected and re-analyzed \chandra, \xmm, \nustar and \bat observations, building a broadband X-ray spectrum ($0.3-150$ keV) and spanning from the early 2000s up to 2016; we have furthermore added results from earlier archival observations taken from the literature, for a total timespan of $\sim 23$ years. Our main findings are as follows:

\begin{itemize}
    \item The soft X-ray emission ($E<2$ keV) is stable across the years (Figure \ref{fig:spec}), and is well described by emission from hot plasma consistent with that found in the anomalous arms on kpc scales.
    
    \item The hard X-ray emission ($E>2$ keV), on the other hand, displays significant variability (Figure \ref{fig:spec}), both on timescales of hours (Figure \ref{fig:lc}) and years (Figure \ref{fig:var}), both intrinsic and due to absorbing gas.
    
    \item We have employed three different models to fit the dataset of 17 spectra: a simple absorbed power law, a torus, and a warped disk. All three models gave consistent results over the spectral parameters, either when fitting the data simultaneously with a single model, or when a full spectral variability is allowed for (Figure \ref{fig:setups}).
    
    \item The average spectral properties of NGC 4258 are typical of low luminosity obscured Seyferts, with a photon index and column density fluctuating in the range $\Gamma = 1.6-2.2$ and $N_{\rm H} = 0.8-2.1 \times 10^{23}$ cm$^{-2}$, respectively (Figure \ref{fig:var}).
    
    \item The obscuring column density shows fluctuations of a factor of two, as previously reported in the literature. Its variations appear to be qualitatively periodic, with a period of about 10 years (Figure \ref{fig:var}). If confirmed, this trend would suggest smooth density variations, possibly induced by spiral density waves and/or gaps in the disk carved by massive He I stars.
    
    \item The Fe K$\alpha$ line, as previously reported in the literature, is weak, with an average $\text{EW} = 45\pm12$ eV. Its detectability is significantly different among the dataset considered here, and in only two epochs it is detected at $\sim 2 \sigma$ level of confidence (Figure \ref{fig:var}).
    
    \item The absorption-corrected, intrinsic $2-10$ keV luminosity is observed to be almost steadily decreasing by a factor of 3 between 2000 and 2016 (Figure \ref{fig:var}).
    
    \item The variations in photon index and luminosity appear to follow two different behaviors (Figure \ref{fig:gammaEddrat}): when the source is brighter than a certain critical value in X-ray-scaled Eddington ratio ($\log{\lambda_{\rm X}} > -5$), no apparent trend is seen; at lower accretion rate, there seems to be an anti-correlation between the two quantities, which may indicate a transition between hot and cold accretion states, similar to what is observed in X-ray binaries.
\end{itemize}

\begin{acknowledgements}
We thank J. Buchner for useful guidance in using the warped disk model. 
This work made use of data from the \nustar
mission, a project led by the California Institute of Technology, managed by the Jet Propulsion Laboratory, and funded by the National Aeronautics and Space Administration. 
This research made use of the \nustar Data Analysis Software (NuSTARDAS) jointly developed by the ASI Science Data Center (ASDC, Italy) and the California Institute of Technology (USA). This research has also made use of data obtained
from the \chandra Data Archive and software provided by the \chandra X-ray Center
(CXC). This work is also based on observations obtained with \xmm, an ESA science mission with instruments and contributions directly funded by ESA Member States and NASA. We acknowledge the use of public data from the {\it Swift} data archive.

P. B. acknowledges financial support from the Czech Science Foundation project No. 22-22643S.
\end{acknowledgements}

\bibliographystyle{aa} 
\bibliography{bib} 

\onecolumn
\begin{appendix}
\section{Additional tables}\label{sec:appendix}

\begin{center}
\begin{longtable}{l c c c} \label{tab:A}
\\
\caption{Results of the analysis with the three models assumed in setup A.} \\

\hline \textbf{Parameter} & \textbf{ABSPL} & \textbf{BORUS} & \textbf{WARPDSK} \\ \hline 
\endfirsthead

\multicolumn{4}{c}
{{\bfseries \tablename\ \thetable{} -- continued}} \\
\hline \textbf{Parameter} & \textbf{ABSPL} & \textbf{BORUS} & \textbf{WARPDSK} \\ \hline 
\endhead

\hline \hline
\endlastfoot

\hline
    $\chi^2$/dof & 3976/3761  & 3969/3760 & 3976/3760\\
\hline
    \multicolumn{4}{c}{Soft X-ray emission} \\
\hline 
    $kT_1$ [keV] & $0.12^{+0.04}_{-0.01}$ & $0.12^{+0.03}_{-0.01}$ & $0.15 \pm 0.04$ \\
    $K_{\rm mekal1}$ [photons keV$^{-1}$ cm$^{-2}$ s$^{-1}$ at 1 keV] & $4.26^{+1.66}_{-2.21} \times 10^{-5}$ & $5.05^{+1.59}_{-2.60} \times 10^{-5}$ & $2.46^{+2.63}_{-0.74} \times 10^{-5}$ \\
    $kT_2$ [keV] & $0.57^{+0.01}_{-0.02}$  & $0.57^{+0.01}_{-0.02}$  & $0.57^{+0.01}_{-0.02}$ \\
    $K_{\rm mekal2}$ [photons keV$^{-1}$ cm$^{-2}$ s$^{-1}$ at 1 keV] & $4.38 \pm 0.17 \times 10^{-5}$ & $4.57 \pm 0.17 \times 10^{-5}$ & $4.30 \pm 0.19 \times 10^{-5}$  \\
    $E_{\rm Si K\alpha}$ [keV] & $1.83 \pm 0.05$ & $1.86 \pm 0.04$ & $1.84^{+0.04}_{-0.06}$\\
    $EW_{\rm Si K\alpha}$ [eV] & $41^{+28}_{-14}$ & $45^{+19}_{-21}$ & $40^{+23}_{-19}$ \\
    $K_{\rm SPL}$ [photons keV$^{-1}$ cm$^{-2}$ s$^{-1}$ at 1 keV] & $4.32 \pm 0.23 \times 10^{-5}$ & $3.80^{+0.23}_{-0.22} \times 10^{-5}$ & $1.18^{+1.69}_{-0.76} \times 10^{-4}$ \\
\hline 
    \multicolumn{4}{c}{Hard X-ray emission} \\
\hline 
    $\Gamma$ & $1.78 \pm 0.03$ & $1.80^{+0.03}_{-0.02}$ & $1.87 \pm 0.03$\\
    $K_{\rm PL}$ [photons keV$^{-1}$ cm$^{-2}$ s$^{-1}$ at 1 keV] & $4.20^{+0.24}_{-0.22} \times 10^{-3}$ & $4.27^{+0.24}_{-0.20} \times 10^{-3}$ & $8.40^{+4.34}_{-4.42}\times 10^{-3}$ \\
    $N_{\rm H}$ [$\times 10^{22}$ cm$^{-2}$] & $9.50 \pm 0.20$ & $9.62 \pm 0.20$ & $ 8.95^{+0.22}_{-0.24}$ \\
    $EW_{\text{Fe K}\alpha}$ [eV] & $45 \pm 12$ & $-$ & $-$ \\
    $\log{N_{\rm H, Tor}}$ & $-$ & $23.43^{+0.15}_{-0.09}$ & $-$ \\
    CF & $-$ & $<0.16$ & $-$ \\
    $\log{N_{\rm H, Disk}}$ & $-$ & $-$ & $25.0^{+0.4}_{-0.3}$ \\
    Disk fraction & $-$ & $-$ & $0.88^{+0.09}_{-0.36}$ \\
\hline
    \multicolumn{4}{c}{Average flux of Epoch 1} \\
\hline 
    $F^{\rm obs}_{2-10}$ [\fluxcgs] & $8.44 \pm 0.21 \times 10^{-12}$ & $8.43^{+0.24}_{-0.31} \times 10^{-12}$ &$8.42^{+0.23}_{-5.07} \times 10^{-12}$ \\
    $F^{\rm int}_{2-10}$ [\fluxcgs] & $1.58 \times10^{-11}$ & $1.56 \times 10^{-11}$ & $2.35 \times 10^{-11}$  \\
\hline
    \multicolumn{4}{c}{Nuclear component relative to \xmm PN, epoch 1} \\
\hline
    
    \xmm MOS, Epoch 1 & $1.13 \pm 0.03$ & $1.13 \pm 0.03$ & $1.13 \pm 0.03$ \\
    \xmm PN, Epoch 2 & $0.84 \pm 0.03$ & $0.83 \pm 0.03$ & $0.84 \pm 0.03$ \\
    \xmm MOS, Epoch 2 & $0.86\pm 0.03$ & $0.86 \pm 0.03$ & $0.86 \pm 0.03$ \\
    \xmm PN, Epoch 3 &$0.74 \pm 0.04$ & $0.73 \pm 0.04$ & $0.74 \pm 0.03$ \\
    \xmm MOS, Epoch 3 &$0.77 \pm 0.03$ & $0.77 \pm 0.03$ & $0.77 \pm 0.03$ \\
    \xmm PN, Epoch 4 &$0.64 \pm 0.13$ &$0.63 \pm 0.13$ & $0.64 \pm 0.13$\\
    \xmm MOS, Epoch 4 &$0.60 \pm 0.04$ & $0.60 \pm 0.05$&  $0.61 \pm 0.04$ \\
    \xmm PN, Epoch 5 &$0.47 \pm 0.02$ & $0.46 \pm 0.02$ & $0.47 \pm 0.02$ \\
    \xmm MOS, Epoch 5 & $0.44 \pm 0.02$ &$0.43 \pm 0.02$ & $0.44 \pm 0.02$ \\
    \xmm PN, Epoch 6 &$0.44 \pm 0.01$ & $0.43 \pm 0.01$ & $0.44 \pm 0.01$ \\
    \xmm MOS, Epoch 6 & $0.48 \pm 0.01$ & $0.46 \pm 0.01$ & $0.48 \pm 0.01$ \\
    \nustar FPMA, Epoch 7 &$0.31 \pm 0.01$ &$0.30 \pm 0.01$ & $0.32 \pm 0.01$ \\
    \nustar FPMB, Epoch 7 &$0.33 \pm 0.01$ &$0.31 \pm 0.01$ & $0.33 \pm 0.01$ \\
    \nustar FPMA, Epoch 8 & $0.25 \pm 0.01$ & $0.23 \pm 0.01$ & $0.25 \pm 0.01$ \\
    \nustar FPMB, Epoch 8 & $0.25 \pm 0.01$ & $0.23 \pm 0.01$ & $0.25 \pm 0.01$ \\
    \bat, Average & $0.66 \pm 0.08$ & $0.67 \pm 0.08$ &  $0.69 \pm 0.09$ \\

\end{longtable}
\end{center}


\begin{center}
\begin{longtable}{l c c c} \label{tab:B} 
\\
\caption{Results of the analysis with the three models assumed in setup B.} \\

\hline \textbf{Parameter} & \textbf{ABSPL} & \textbf{BORUS} & \textbf{WARPDSK} \\ \hline 
\endfirsthead

\multicolumn{4}{c}
{{\bfseries \tablename\ \thetable{} -- continued}} \\
\hline \textbf{Parameter} & \textbf{ABSPL} & \textbf{BORUS} & \textbf{WARPDSK} \\ \hline 
\endhead

\hline \hline
\endlastfoot

\hline
    $\chi^2$/dof & 3620/3732 & 3624/3739 & 3653/3737 \\
\hline
    \multicolumn{4}{c}{Soft X-ray emission} \\
\hline \hline
    $kT_1$ [keV] & $0.11 \pm 0.01$  & $0.11 \pm 0.01$ & $0.11 \pm 0.01$ \\
    $K_{\rm mekal1}$ [photons keV$^{-1}$ cm$^{-2}$ s$^{-1}$ at 1 keV] & $8.18^{+2.89}_{-1.64} \times 10^{-5}$ & $8.38^{+2.85}_{-1.74} \times 10^{-5}$ & $7.54^{+1.28}_{-0.57} \times 10^{-5}$ \\
    $kT_2$ [keV] & $0.54 \pm 0.01$ & $0.54 \pm 0.02$ & $0.54 \pm 0.02$ \\
    $K_{\rm mekal2}$ [photons keV$^{-1}$ cm$^{-2}$ s$^{-1}$ at 1 keV] & $4.63 \pm 0.16 \times 10^{-5}$ & $4.68^{+0.15}_{-0.18} \times 10^{-5}$ & $4.60 \pm 0.12 \times 10^{-5}$ \\
    $E_{\rm Si K\alpha}$ [keV] & $1.85^{+0.03}_{-0.07}$ & $1.84^{+0.04}_{-0.06}$ & $1.85^{+0.04}_{-0.06}$\\
    $EW_{\rm Si K\alpha}$ [eV] & $32^{+15}_{-17}$ & $29^{+18}_{-14}$ & $28^{+17}_{-16}$ \\
    CF & $-$ & $0.37 \pm 0.14$ & $-$ \\
    Disk fraction & $-$ & $-$ & $0.25 \pm 0.01$\\
\hline 
    \multicolumn{4}{c}{Hard X-ray emission} \\
\hline \hline
    \multicolumn{4}{c}{Epoch 1 - \xmm PN/MOS} \\ 
    $\Gamma$ & $1.67 \pm 0.08$ & $1.67^{+0.09}_{-0.07}$ & $1.80^{+0.03}_{-0.01}$ \\
    $K_{\rm PL}$ [photons keV$^{-1}$ cm$^{-2}$ s$^{-1}$ at 1 keV] & $3.27^{+0.53}_{-0.45} \times 10^{-3}$ & $3.27^{+0.62}_{-0.39} \times 10^{-3}$ & $4.87^{+0.11}_{-0.34} \times 10^{-3}$\\
    $N_{\rm H}$ [$\times 10^{22}$ cm$^{-2}$] & $8.33 \pm 0.38$ & $8.32^{+0.59}_{-0.19}$ & $8.47 \pm 0.14$\\
    $K_{\rm SPL}$ [photons keV$^{-1}$ cm$^{-2}$ s$^{-1}$ at 1 keV] & $2.40 \pm 0.38 \times 10^{-5}$  & $2.27^{+0.43}_{-0.35} \times 10^{-5}$ & $5.17^{+0.63}_{-0.66} \times 10^{-5}$\\
    $EW_{\text{Fe K}\alpha}$ [eV] & $14^{+22}_{-14}$ & $-$ & $-$ \\
    $F_{\text{Fe K}\alpha}$ [\fluxcgs] & $1.70^{+2.77}_{-l} \times 10^{-14}$ & $-$ & $-$ \\
    MOS/PN & $1.11 \pm 0.03$ & $1.11 \pm 0.03$ & $1.11 \pm 0.03$ \\
    $F^{\rm obs}_{2-10}$ [\fluxcgs] & $8.42^{+0.11}_{-0.16} \times 10^{-12}$ & $8.41^{+0.10}_{-0.17} \times 10^{-12}$ & $8.33^{+0.23}_{-0.03} \times 10^{-12}$ \\
    $F^{\rm int}_{2-10}$ [\fluxcgs] & $\sim 1.46 \times 10^{-11}$ & $\sim 1.44 \times 10^{-11}$ & $\sim 1.48 \times 10^{-11}$ \\
    \hline
    
    \multicolumn{4}{c}{Epoch 2 - \xmm PN/MOS} \\ 
    $\Gamma$ & $1.85 \pm 0.12$ & $1.86^{+0.13}_{-0.11}$ & $1.97^{+0.01}_{-0.02}$ \\
    $K_{\rm PL}$ [photons keV$^{-1}$ cm$^{-2}$ s$^{-1}$ at 1 keV] & $4.06^{+1.05}_{-0.83} \times 10^{-3}$ & $4.08^{+1.11}_{-0.80} \times 10^{-3}$ & $5.85^{+0.65}_{-0.61} \times 10^{-3}$ \\
    $N_{\rm H}$ [$\times 10^{22}$ cm$^{-2}$] & $9.93^{+0.63}_{-0.61}$ & $10.0^{+0.7}_{-0.7}$ & $9.98^{+0.23}_{-0.22}$\\
    $K_{\rm SPL}$ [photons keV$^{-1}$ cm$^{-2}$ s$^{-1}$ at 1 keV] & $3.17 \pm 0.49 \times 10^{-5}$  & $3.05^{+0.52}_{-0.47} \times 10^{-5}$ & $6.77^{+0.86}_{-0.89} \times 10^{-5} $\\
    $EW_{\text{Fe K}\alpha}$ [eV] & $32^{+35}_{-32}$ & $-$ & $-$ \\
    $F_{\text{Fe K}\alpha}$ [\fluxcgs] & $3.39^{+3.69}_{-l} \times 10^{-14}$ & $-$ & $-$ \\
    MOS/PN & $1.03 \pm 0.04$ & $1.03 \pm 0.04$ & $1.03 \pm 0.04$ \\
    $F^{\rm obs}_{2-10}$ [\fluxcgs] & $6.74^{+0.10}_{-0.21} \times 10^{-12}$ & $6.74^{+0.12}_{-0.19} \times 10^{-12}$ & $6.69^{+0.23}_{-0.13} \times 10^{-12}$\\
    $F^{\rm int}_{2-10}$ [\fluxcgs] & $\sim 1.31 \times 10^{-11}$ & $\sim 1.30 \times 10^{-11}$ & $\sim 1.33 \times 10^{-11}$\\
    \hline
    
    \multicolumn{4}{c}{Epoch 3 - \xmm PN/MOS} \\ 
    $\Gamma$ & $1.98 \pm 0.18$ & $1.96^{+0.21}_{-0.16}$ & $2.09^{+0.04}_{-0.02}$\\
    $K_{\rm PL}$ [photons keV$^{-1}$ cm$^{-2}$ s$^{-1}$ at 1 keV] & $4.24^{+1.75}_{-1.22} \times 10^{-3}$ & $4.10^{+1.94}_{-1.04} \times 10^{-3}$ & $6.07^{+0.99}_{-0.92} \times 10^{-3}$\\
    $N_{\rm H}$ [$\times 10^{22}$ cm$^{-2}$] & $9.11^{+0.85}_{-0.81}$ & $9.12^{+0.88}_{-0.61}$ & $9.20^{+0.30}_{-0.28}$\\
    $K_{\rm SPL}$ [photons keV$^{-1}$ cm$^{-2}$ s$^{-1}$ at 1 keV] & $2.40 \pm 0.70 \times 10^{-5}$  & $2.27^{+0.77}_{-0.66} \times 10^{-5}$ & $5.07^{+1.22}_{-1.25} \times 10^{-5}$\\
    $EW_{\text{Fe K}\alpha}$ [eV] & $52^{+59}_{-52}$ & $-$ & $-$\\
    $F_{\text{Fe K}\alpha}$ [\fluxcgs] & $4.47^{+5.08}_{-l} \times 10^{-14}$ & $-$ & $-$ \\
    MOS/PN & $1.00 \pm 0.05$ & $1.00^{+0.06}_{-0.05}$ & $1.00^{+0.06}_{-0.05}$ \\
    $F^{\rm obs}_{2-10}$ [\fluxcgs] & $5.80^{+0.17}_{-0.31} \times 10^{-12}$ & $5.81^{+0.14}_{-0.33} \times 10^{-12}$ & $5.78^{+0.15}_{-0.42} \times 10^{-12}$ \\
    $F^{\rm int}_{2-10}$ [\fluxcgs] & $\sim 1.11 \times 10^{-11}$ & $\sim 1.10 \times 10^{-11}$ & $\sim 1.13 \times 10^{-11}$\\
    \hline
    
    \multicolumn{4}{c}{Epoch 4 - \xmm PN/MOS} \\ 
    $\Gamma$ & $2.07^{+0.45}_{-0.43}$ & $2.06^{+0.51}_{-0.40}$ & $2.25^{+0.04}_{-0.12}$ \\
    $K_{\rm PL}$ [photons keV$^{-1}$ cm$^{-2}$ s$^{-1}$ at 1 keV] & $5.46^{+7.66}_{-3.18} \times 10^{-3}$ & $5.18^{+8.62}_{-2.84} \times 10^{-3}$ & $8.67^{+3.74}_{-3.39} \times 10^{-3}$\\
    $N_{\rm H}$ [$\times 10^{22}$ cm$^{-2}$] & $13.1^{+2.5}_{-2.4} $ & $12.9^{+3.0}_{-2.2}$ & $13.2^{+0.8}_{-0.6}$\\
    $K_{\rm SPL}$ [photons keV$^{-1}$ cm$^{-2}$ s$^{-1}$ at 1 keV] & $4.77^{+2.30}_{-2.00} \times 10^{-5}$ & $4.53^{+2.43}_{-1.92} \times 10^{-5}$ & $1.03^{+0.28}_{-0.29} \times 10^{-4}$ \\
    $EW_{\text{Fe K}\alpha}$ [eV] & $< 97$ & $-$ & $-$ \\
    $F_{\text{Fe K}\alpha}$ [\fluxcgs] & $-$ & $-$ & $-$ \\
    MOS/PN & $0.96^{+0.17}_{-0.14}$ & $0.96^{+0.17}_{-0.14}$ & $0.96^{+0.17}_{-0.14}$\\
    $F^{\rm obs}_{2-10}$ [\fluxcgs] & $5.16^{+0.38}_{-2.35} \times 10^{-12}$ & $5.10^{+0.31}_{-1.90} \times 10^{-12}$ & $5.03^{+0.30}_{-1.81} \times 10^{-12}$\\
    $F^{\rm int}_{2-10}$ [\fluxcgs] & $\sim 1.22 \times 10^{-11}$ & $\sim 1.18 \times 10^{-11}$ & $\sim 1.25 \times 10^{-11}$\\
    \hline
    
    \multicolumn{4}{c}{Epoch 5 - \xmm PN/MOS} \\ 
    $\Gamma$ & $1.60 \pm 0.15$ & $1.60^{+0.16}_{-0.15}$ & $1.72^{+0.01}_{-0.03}$\\
    $K_{\rm PL}$ [photons keV$^{-1}$ cm$^{-2}$ s$^{-1}$ at 1 keV] & $2.03^{+0.75}_{-0.55} \times 10^{-3}$ & $2.02^{+0.81}_{-0.52} \times 10^{-3}$ & $2.92^{+0.40}_{-0.38} \times 10^{-3}$\\
    $N_{\rm H}$ [$\times 10^{22}$ cm$^{-2}$] & $14.2 \pm 1.2$ & $14.1^{+1.4}_{-0.9}$ & $14.0 \pm 0.4$\\
    $K_{\rm SPL}$ [photons keV$^{-1}$ cm$^{-2}$ s$^{-1}$ at 1 keV] & $2.64 \pm 0.42 \times 10^{-5}$  & $2.49^{+0.46}_{-0.40} \times 10{-5}$ & $5.64^{+0.70}_{-4.90} \times 10^{-5}$\\
    $EW_{\text{Fe K}\alpha}$ [eV] & $40 \pm 38$ & $-$ & $-$\\
    $F_{\text{Fe K}\alpha}$ [\fluxcgs] & $3.02 \pm 2.87 \times 10^{-14}$ & $-$ & $-$ \\
    MOS/PN & $1.01 \pm 0.04$ & $1.01 \pm 0.04$ & $1.01 \pm 0.04$\\
    $F^{\rm obs}_{2-10}$ [\fluxcgs] & $4.50^{+0.09}_{-0.21} \times 10^{-12}$ & $4.50^{+0.09}_{-0.22} \times 10^{-12}$ & $4.47^{+0.08}_{-0.21} \times 10^{-12}$\\
    $F^{\rm int}_{2-10}$ [\fluxcgs] & $\sim 9.63 \times 10^{-12}$ & $\sim 9.50 \times 10^{-12}$ & $\sim 9.72 \times 10^{-12}$\\
    \hline
    
    \multicolumn{4}{c}{Epoch 6 - \xmm PN/MOS} \\ 
    $\Gamma$ & $1.71 \pm 0.07$ & $1.71^{+0.08}_{-0.06}$ & $1.84^{+0.04}_{-0.01}$ \\
    $K_{\rm PL}$ [photons keV$^{-1}$ cm$^{-2}$ s$^{-1}$ at 1 keV] & $1.58^{+0.23}_{-0.20} \times 10^{-3}$ & $1.60^{+0.26}_{-0.18} \times 10^{-3}$ & $2.35^{+0.15}_{-0.14} \times 10^{-3}$\\
    $N_{\rm H}$ [$\times 10^{22}$ cm$^{-2}$] & $8.59^{+0.38}_{-0.37}$ & $8.71^{+0.41}_{-0.39}$ & $8.75^{+0.14}_{-0.13}$ \\
    $K_{\rm SPL}$ [photons keV$^{-1}$ cm$^{-2}$ s$^{-1}$ at 1 keV] & $2.40 \pm 0.26 \times 10^{-5}$ & $2.28^{+0.29}_{-0.24} \times 10^{-5}$ & $5.17^{+0.35}_{-0.39} \times 10^{-5}$\\
    $EW_{\text{Fe K}\alpha}$ [eV] & $43^{+21}_{-20}$ & $-$ & $-$ \\
    $F_{\text{Fe K}\alpha}$ [\fluxcgs] & $2.40^{+1.15}_{-1.17} \times 10^{-14}$ & $-$ & $-$ \\
    MOS/PN & $1.06 \pm 0.02$ & $1.06 \pm 0.02$ & $1.06 \pm 0.02$\\
    $F^{\rm obs}_{2-10}$ [\fluxcgs] & $3.78^{+0.04}_{-0.05} \times 10^{-12}$ & $3.78^{+0.05}_{-0.05} \times 10^{-12}$ & $3.75 \pm 0.08 \times 10^{-12}$\\
    $F^{\rm int}_{2-10}$ [\fluxcgs] & $\sim 6.49 \times 10^{-12}$ & $\sim 6.44 \times 10^{-12}$ & $\sim 6.61 \times 10^{-12}$\\
    \hline
    
    \multicolumn{4}{c}{Epoch 7 - \nustar FPMA/B} \\ 
    $\Gamma$ & $1.88 \pm 0.13$ & $1.89^{+0.11}_{-0.12}$ & $1.94^{+0.01}_{-0.02}$ \\
    $K_{\rm PL}$ [photons keV$^{-1}$ cm$^{-2}$ s$^{-1}$ at 1 keV] & $1.55^{+0.91}_{-0.46} \times 10^{-3}$ & $1.58^{+0.76}_{-0.44} \times 10^{-3}$ & $2.07^{+0.23}_{-0.18} \times 10^{-3}$\\
    $N_{\rm H}$ [$\times 10^{22}$ cm$^{-2}$] & $20.1^{+9.8}_{-10.7}$ & $20.9^{+8.6}_{-9.9}$ & $11.1^{+0.6}_{-0.9}$\\
    $\log{N_{\rm H, Disk}}$ & $-$ & $-$ & $24^{+0.4}_{-l}$\\
    $K_{\rm SPL}$ [photons keV$^{-1}$ cm$^{-2}$ s$^{-1}$ at 1 keV] & $4.0^{+2.1}_{-3.8} \times 10^{-4}$ & $4.17^{+1.88}_{-3.20} \times 10^{-4}$ & $2.07^{+u}_{-l} \times 10^{-4}$**\\
    $EW_{\text{Fe K}\alpha}$ [eV] & $22^{+52}_{-22}$ & $-$ & $-$ \\
    $F_{\text{Fe K}\alpha}$ [\fluxcgs] & $0.91^{+2.18}_{-l} \times 10^{-14}$ & $-$ & $-$ \\
    FPMB/FPMA & $1.03 \pm 0.04$ & $1.03 \pm 0.04$ & $1.03 \pm 0.04$\\
    $F^{\rm obs}_{2-10}$ [\fluxcgs] & $2.89^{+0.10}_{-0.17} \times 10^{-12}$ & $2.90^{+0.04}_{-0.33} \times 10^{-12}$ & $2.72^{+1.55}_{-0.72} \times 10^{-12}$\\
    $F^{\rm int}_{2-10}$ [\fluxcgs] & $\sim 4.77 \times 10^{-12}$ & $\sim 4.76 \times 10^{-12}$ & $\sim 4.95 \times 10^{-12}$\\
    \hline
    
    \multicolumn{4}{c}{Epoch 8 - \nustar FPMA/B} \\ 
    $\Gamma$ & $1.92^{+0.12}_{-0.11}$ &  $2.00^{+0.09}_{-0.12}$ & $2.05^{+0.01}_{-0.02}$\\
    $K_{\rm PL}$ [photons keV$^{-1}$ cm$^{-2}$ s$^{-1}$ at 1 keV] & $1.44^{+0.57}_{-0.28} \times 10^{-3}$ & $1.74^{+0.58}_{-0.42} \times 10^{-3}$ & $2.23^{+0.21}_{-0.14} \times 10^{-3}$\\
    $N_{\rm H}$ [$\times 10^{22}$ cm$^{-2}$] & $14.6^{+8.5}_{-5.7}$ & $19.5^{+6.8}_{-8.8}$ & $12.8^{+0.5}_{-1.2}$\\
    $\log{N_{\rm H, Disk}}$ & $-$ & $-$ & $24^{+0.3}_{-l}$\\
    $K_{\rm SPL}$ [photons keV$^{-1}$ cm$^{-2}$ s$^{-1}$ at 1 keV] & $1.80^{+2.21}_{-1.80} \times 10^{-4}$ & $3.09^{+1.52}_{-3.09} \times 10^{-4}$ & $2.23^{+u}_{-l} \times 10^{-4}$**\\
    $EW_{\text{Fe K}\alpha}$ [eV] & $65^{+46}_{-45}$ & $-$ & $-$\\
    $F_{\text{Fe K}\alpha}$ [\fluxcgs] & $2.24 \pm 1.56 \times 10^{-14}$ & $-$ & $-$ \\
    FPMB/FPMA & $0.98 \pm 0.03$ & $0.98 \pm 0.03$ & $0.98 \pm 0.03$\\
    $F^{\rm obs}_{2-10}$ [\fluxcgs] & $2.24^{+0.18}_{-0.06} \times 10^{-12}$ & $2.29^{+0.05}_{-0.12} \times 10^{-12}$ & $2.19^{+0.31}_{-0.46} \times 10^{-12}$\\
    $F^{\rm int}_{2-10}$ [\fluxcgs] & $\sim 4.05 \times 10^{-12}$ & $\sim 4.36 \times 10^{-12}$ & $\sim 4.37 \times 10^{-12}$ \\
    \hline
    
    \multicolumn{4}{c}{\bat} \\ 
    $\Gamma$ & $1.67^{+0.23}_{-0.22}$ & $1.67^{+0.23}_{-0.22}$ & $1.73^{+0.03}_{-0.18}$ \\
    $K_{\rm PL}$ [photons keV$^{-1}$ cm$^{-2}$ s$^{-1}$ at 1 keV] & $1.91^{+2.44}_{-1.08} \times 10^{-3}$ & $1.90^{+2.38}_{-1.07} \times 10^{-3}$ & $2.20^{+0.25}_{-1.32} \times 10^{-3}$\\
    $F^{\rm obs}_{14-195}$ [\fluxcgs] & $2.21^{+0.18}_{-0.92} \times 10^{-11}$ & $2.21^{+0.23}_{-0.79} \times 10^{-11}$ & $2.16^{+0.41}_{-1.00} \times 10^{-11}$\\
    $F^{\rm int}_{14-195}$ [\fluxcgs] & $\sim 2.27 \times 10^{-11}$ & $\sim 2.24 \times 10^{-11}$ & $\sim 2.24 \times 10^{-11}$\\
\hline
\end{longtable}
\end{center}

\end{appendix}
\end{document}